\documentstyle[prc,aps,epsfig]{revtex}

\renewcommand{\l}{\left} 
\newcommand{\r}{\right}

\newcommand{\N}{{\cal N}}

\newcommand{\dd}{{\rm d}}
\newcommand{\pd}{\partial}
\begin{document}


\draft

\twocolumn[
\hsize\textwidth\columnwidth\hsize\csname @twocolumnfalse\endcsname

\title{Equation of State for the Two-component Van der Waals Gas \\
       with Relativistic Excluded Volumes}

\author{G. Zeeb\footnotemark[1] $^{\dagger}$,
  K.\,A. Bugaev$^{\ddagger\,**}$, P.\,T. Reuter$^{\dagger}$
  and  H. St\"ocker$^{\dagger}$}

\address{$^{\dagger}$ Institut f\"ur Theoretische Physik,
  J.-W.-Goethe Universit\"at  Frankfurt am Main, Germany  \\ 
$^{\ddagger}$ Gesellschaft f\"ur Schwerionenforschung mbH (GSI),
  Darmstadt, Germany  \\
$^{**}$ Bogolyubov Institute for Theoretical Physics,
  252143 Kiev, Ukraine}

\maketitle

\begin{abstract}
  A canonical partition function for the two-component excluded volume model
  is derived, leading to two different {\sl van der Waals\/} approximations.
  The one is known as the {\sl Lorentz-Berthelot\/} mixture and the other
  has been proposed recently.
  Both models are analysed in the canonical and grand canonical ensemble.
  In comparison with the one-component {\sl van der Waals\/} excluded
  volume model the suppression of particle densities is reduced in these
  two-component formulations, but in two essentially different ways.
  Presently used multi-component models have no such reduction.
  They are shown to be not correct when used for components with
  different hard-core radii.

  For high temperatures the excluded volume interaction is refined by
  accounting for the {\sl Lorentz\/} contraction of the spherical excluded
  volumes, which leads to a distinct enhancement of lighter particles.
  The resulting effects on pion yield ratios are studied for AGS and SPS
  data.
\end{abstract}


\begin{quote}
  KEYWORDS\,:\,
  Two-component and Multi-component Hadron Gas, Equation of State,  \\
  Van der Waals Excluded Volume Model, Relativistic Excluded Volumes\,. \\
\end{quote}
]

\narrowtext

\footnotetext[1]{e-mail: gzeeb\symbol{64}th.physik.uni-frankfurt.de}


\section{Introduction}

Thermal models are commonly used to interprete experimental data
from hadron collisions, see for instance
\cite{stoecker,rafelski,cleymans,braun-munzinger,becattini}\,.
In the {\sl van der Waals\/} excluded volume {\sl (VdW)\/} model the
short range repulsion between particles is represented by hard-core
potentials, i.\,e.~the finite size of the particles is taken into account.
As a consequence particle yields are essentially reduced in comparison
with ideal gas results, whereas yield ratios remain almost unchanged,
if the same hard-core radius is used for all particle species.

As particle species with smaller hard-core radii are closer to the
ideal case, their particle densities are suppressed less.
Consequently, their yield ratios to particle species with larger
hard-core radii are enhanced.
This fact has been used in recent efforts \cite{Yen:97} to explain the
experimentally observed pion abundance for AGS and SPS data \cite{PrQM:96}
by introducing a smaller hard-core radius for pions $R_\pi$ than for
all other hadrons $R_{\rm o}$\,.
However, the resulting values are quite large, $R_{\rm o}=0.8$ fm and
$R_\pi=0.62$ fm\,.
In Ref.~\cite{BrM:99} a reasonable fit of SPS data has been obtained only
for a distinctly smaller pair of hard-core radii.

The excluded volume models used in \cite{Yen:97,BrM:99}\,, 
however, are not correct in the case of different hard-core radii.
As will be shown in Sect.~II\,, these models correspond to a system where
the components are separated from each other by a mobile wall and hence
cannot mix.

A more realistic approach requires a two-component partition function
including a term for the repulsion between particles of different hard-core
radii.
In the case of two components, however, the {\sl VdW\/} approximation is not
uniquely defined.
The simplest possibility yields the {\sl Lorentz-Berthelot\/} mixture,
which was originally postulated by {\sl van der Waals\/} for binary mixtures,
see Refs.~\cite{VdW:1889,Lor:1881,Berth:1898,Muen}\,.
Another {\sl VdW\/} approximation was recently proposed in Ref.~\cite{Gor:99}\,.
These two formulations contain a suppression of particle densities
similar to the one-component {\sl van der Waals\/} gas, which is
{\em reduced to different extend\/} for each formulation.
In the present work we will study and apply both of these formulations.

There is yet another cause for a reduced excluded volume suppression.
Particles are considered to be rigid spheres in the {\sl VdW\/} model.
At high energies as achieved in nuclear collisions, however, relativistic
effects cannot be neglected \cite{Bug:00}\,.
Within the logic of the {\sl VdW\/} model it is necessary to take into
account the {\sl Lorentz\/} contraction of the spheres.
We will use an approach developed in Ref.~\cite{Bug:99} providing
approximative formulae for relativistic excluded volumes:
naturally, they decrease with rising temperature, and the effect is
stronger for lighter particles.
At high temperatures, consequently, it is not possible to use a
{\em one-component\/} {\sl VdW\/} description (i.\,e.~a {\em common\/}
excluded volume for {\em all\/} particle species) for a system of
species with various masses.
Since different masses cause different reductions of the excluded volumes at
a given temperature, a {\em multi-component\/} {\sl VdW\/} description is required.

To illustrate the influence of different excluded volumes we will
restrict ourselves in this work to the simplest 'multi-component' case,
the two-component case.
The crucial extension from the one- to the two-component case is to include
the repulsion between particles of two {\em different\/} hard-core radii.
As it will be illustrated, a generalisation to the multi-component case
is straightforward and will yield no essential differences \cite{Zeeb:02}\,.
%

In the next section a derivation of the one-component canonical
partition function with (constant) excluded volumes is presented.
The generalisation to the two-component case is made and two possible
{\sl VdW\/} approximations are analysed:
the {\sl Lorentz-Berthelot\/} mixture \cite{Muen}
and the recently proposed approximation of Ref.~\cite{Gor:99}\,.
The corresponding formulae for the grand canonical ensemble are derived
and discussed in Sect.~III\,.
Relativistic excluded volumes are introduced in Sect.~IV and
the corresponding equations of state are presented for both models.
In Sect.~V a fit of particle yield ratios \cite{Yen:97} is re-evaluated for
both approximations, with constant and with relativistic excluded volumes.
The conclusions are given in Sect.~VI\,.

The appendixes \ref{CE_appd} and \ref{GCE_appd} give a detailed analysis
and comparison of the two approximations in the canonical and grand
canonical ensemble, respectively.
In App.\ \ref{nl_appd} a general proof of the thermodynamical stability
of the non-linear approximation is given.


\section{Canonical Treatment}

First we will derive the canonical partition function (CPF) for the
one-component {\sl VdW\/} gas by estimating the excluded volumes of
particle clusters.
Then this procedure will be generalised to the two-component case.

For simplicity {\sl Boltzmann\/} statistics are used throughout this work.
The deviations from quantum statistics are negligible as long as the density
over temperature ratio is small.
This is the case for the hadron gas at temperatures and densities typical
for heavy ion collisions, see e.\,g.~Ref.~\cite{Yen:97}\,. 

Note that in this work we will use the term {\sl VdW\/} for the
{\sl van der Waals\/} excluded volume model, not for the general
{\sl van der Waals\/} model which includes attraction.

\subsection{The Van der Waals Excluded Volume Model}

Let us consider $N$ identical particles with temperature $T$ kept in
a sufficiently large volume $V$\,, so that finite volume effects can be
neglected.
The partition function of this system ($\hbar = c = k_{\rm B} = 1$) reads
\begin{eqnarray}
\label{eq:Z-CE}
  Z(T,V,N) &=& \frac{\phi^N}{N!}
               \int_{V^N} \dd^3 x_{1}
                              ~\cdots~ \dd^3 x_N
                              \,\exp \l[{\textstyle -\frac{U_N}{T} }\r] ~.
\end{eqnarray}
Here, $\phi \equiv \phi(T;m,g)$ denotes the momentum integral of the
one particle partition
\begin{equation}
  \phi(T;m,g) = \frac{g}{2 \pi^2} \int\limits_0^{\infty} \dd k \, k^2
                ~\exp\l[{\textstyle - \frac{E(k)}{T} }\r] ~,
\end{equation}
where $E(k) \equiv \sqrt{k^2 + m^2}$ is the relativistic energy and
$g=(2S+1)(2I+1)$ counts the spin and isospin degeneracy.
For a hard-core potential $U_N$ of $N$ spherical particles with radii $R$
the potential term in Eq.~(\ref{eq:Z-CE}) reads
\begin{equation}
  \label{eq:pot1c}
  \exp\l[{\textstyle -\frac{U_N}{T} }\r]
    = {\textstyle \prod\limits_{i<j\le N} } \theta(\l|\vec{x}_{ij}\r| - 2R)~,
\end{equation} 
where $\vec{x}_{ij}$ denotes the relative position vector connecting the
centers of the $i$-th and $j$-th particle.
Hence one can write
\begin{eqnarray}
  \label{eq:Intgrls}
  \lefteqn{\int_{V^N} \dd^3 x_1 ~\cdots~ \dd^3 x_N
           \,\exp\l[{\textstyle -\frac{U_N}{T} }\r]}  \nonumber \\
  &\quad =&
    \int_{V^N} \dd^3 x_1 ~\cdots~ \dd^3 x_N
    {\textstyle \prod\limits_{1 \le i<j \le N} }
         \!\!\!\! \theta (\l| \vec x_{ij} \r| - 2R)  \nonumber \\
  &\quad =&
    \int_V \dd^3 x_1 ~
    \int_V \dd^3 x_2 ~ \theta (\l| \vec x_{12} \r| - 2R) ~\times  \nonumber \\
  &  &
    ~\times~\cdots~ \int_V \dd^3 x_N
    {\textstyle \prod\limits_{1 \le i \le N-1} }
         \!\! \theta (\l| \vec x_{i, N} \r| - 2R)  \nonumber \\
  &\quad \equiv&
    \int_V \dd^3 x_1 \int\limits^{\{ \vec x_1 \}}{\dd^3 x_2}
    ~\cdots \int\limits^{\{\vec x_1 \dots \vec x_{N-1}\}} \!\dd^3 x_N ~.
\end{eqnarray}
Here, $\int\limits^{\{\vec x_1 \dots \vec x_{j}\}} \!\dd^3 x_{j+1}$
denotes the available volume for $\vec x_{j+1}$, which is the center of the
particle with number $j+1$\,, if the $j$ other particles are configurated
as $\{\vec x_1 \dots \vec x_{j}\}$\,.
We will show now that this volume is estimated by
$\int\limits^{\{\vec x_1 \dots \vec x_j\}} \!\dd^3 x_{j+1} \ge (V - 2b\,j)$\,,
where $2b \equiv \frac{4 \pi}{3} (2R)^3$ is the excluded volume of an isolated
particle seen by a second one.
Then, $2b\,j$ estimates the total volume which is excluded by all particle
clusters occuring in the configuration $\{\vec x_1 \dots \vec x_j\}$\,.

It is sufficient to prove that the excluded volume of a cluster of $k$
particles is less than the excluded volume of $k$ isolated particles.
A group of $k$ particles forms a $k$-cluster, if for any of these particles
there is a neighbouring particle of this group at a distance less than $4R$\,.
The {\em exact\/} excluded volume of a $k$-cluster, $v_{(k)}$\,, obviously
depends on the configuration of the $k$ particles.
If one considers two isolated particles, i.\,e.~two $1$-clusters, and reduces
their distance below $4R$\,, their excluded volumes will overlap.
They form now a $2$-cluster with the excluded volume
$v_{(2)}=4b-1 v_{\rm ov}$\,,
where $v_{\rm ov}$ denotes the overlap volume.

Evidently, one can construct any $k$-cluster by attaching additional
particles and calculate its excluded volume by subtracting each occuring 
overlap volume from $2b\,k$\,.
It follows that $v_{(k)} < 2b\,k$ is valid for any $k$-cluster, and this
inequality leads to the above estimate.
Obviously, its accuracy improves with the diluteness of the gas.

Using these considerations one can approximate the r.\,h.\,s.~of
Eq.~(\ref{eq:Intgrls})\,:
starting with $j+1=N$ one gradually replaces all integrals
$\int\limits^{\{ \vec x_1 \dots \vec x_{j} \} } {\!\dd^3 x_{j+1} }$
 by $(V - 2b\,j)$\,.
One has to proceed from the right to the left, because only the respective
rightmost of these integrals can be estimated in the described way.
Hence one finds
\begin{eqnarray}
  \label{eq:Zprox}
  Z(T,V,N)
    & \ge & \frac{\phi^N}{N!}\,
            ~ {\textstyle \prod\limits_{j=0}^{N-1} (V - 2b\,j) }~.
\end{eqnarray}

In this treatment the {\sl VdW\/} approximation consists of two assumptions
concerning Eq.~(\ref{eq:Zprox})\,.
Firstly, the product can be approximated by
\begin{eqnarray}
  \lefteqn{ {\textstyle \prod\limits_{j=0}^{N-1} \l( 1 - \frac{2b}{V}\,j \r)}
      \cong \exp\l[ {\textstyle - \sum_{j=0}^{N-1} \frac{2b}{V}\,j} \r]}  \\
  &\quad =&
      \exp \l[ {\textstyle - \frac{b}{V}\,(N-1)N }\r]
      \cong \l( {\textstyle 1 - \frac{b}{V}\,N }\r)^N ~,  \nonumber
\end{eqnarray}
where $\exp\,[-x] \cong (1-x)$ is used for dilute systems,
i.\,e.\ for low densities $2bN/V \ll 1$\,.
The second assumption is to take the equality
instead of the inequality in Eq.~(\ref{eq:Zprox})\,.
Then the CPF takes the {\sl VdW\/} form,
\begin{equation}
  \label{eq:Z_VdW}
  Z_{\rm VdW}(T,V,N) = \frac{\phi^N}{N!} \,\l( V - bN \r)^N ~.
\end{equation}
As usual, the {\sl VdW\/} CPF is obtained as an approximation for dilute
systems, but when used for high densities it should be considered as
an {\em extrapolation\/}.

Finally, one obtains the well-known {\sl VdW\/} pressure formula from the
thermodynamical identity $p(T,V,N) \equiv T\,\pd \ln[Z(T,V,N)]/\pd V$\,,
\begin{equation}
  \label{eq:p_VdW}
  p_{\rm\,VdW}(T,V,N) = \frac{T \, N}{V - b N} ~,
\end{equation}
using  the logarithm of the {\sl Stirling\/} formula. 

Now let us brief\/ly investigate a system of volume $V$
containing two components with {\em different\/} hard-core radii $R_1$ and
$R_2$ which are separated by a wall and occupy the volume fractions $x V$
and $(1-x) V$\,, respectively.
According to Eq.~(\ref{eq:p_VdW}) their pressures read
\begin{eqnarray}
  \label{eq:p-sp_x}
  p_{\rm\,VdW}(T,xV,N_1)   &=& \frac{T \, N_1}{x V - N_1 b_{11}} ~,  \\
  \label{eq:p-sp_1-x}
  p_{\rm\,VdW}(T,(1-x)V,N_2) &=& \frac{T \, N_2}{(1-x) V - N_2 b_{22}} ~,
\end{eqnarray}
where the particle numbers $N_1, N_2$ and the excluded volumes
$b_{11}=\frac{16 \pi}{3}\,R_1^{\,3}\,,\ b_{22}=\frac{16 \pi}{3}\,R_2^{\,3}$
correspond to the components 1 and 2\,, respectively.

If the separating wall is mobile, the pressures (\ref{eq:p-sp_x})
and (\ref{eq:p-sp_1-x}) must be equal.
In this case the fraction $x$ can be eliminated and
one obtains the common pressure of the whole system
\begin{eqnarray}
  & & p_{\rm\,VdW}(T,xV,N_1) = p_{\rm\,VdW}(T,(1-x)V,N_2)  \nonumber \\
  \label{eq:p-sp_CE}
  &=~& p^{\rm\,sp}(T,V,N_1,N_2)
      \equiv \frac{T\,(N_1+N_2)}{V - N_1 b_{11} - N_2 b_{22}} ~.
\end{eqnarray}
Since the components are separated in this model system
it will be referred to as the {\em separated\/} model \cite{Zeeb:02}\,.

The pressure formula (\ref{eq:p-sp_CE}) corresponds to the
{\sl Boltzmann\/} approximation of the commonly used two-component
{\sl VdW\/} models of Refs.\ \cite{Yen:97,BrM:99} 
as will be shown in Sect.~V\,.
It is evident that $p^{\rm\,sp}$ (\ref{eq:p-sp_CE}) does not
describe the general two-component situation without a separating wall.
Therefore, it is necessary to find a more realistic model, i.\,e.~an
approximation from a {\em two-component\/} partition function.
This will be done in the following.

\subsection{Generalisation to the Two-component Case}

Recall the simple estimate (\ref{eq:Intgrls}--\ref{eq:Z_VdW})\,,
which gives a physically transparent derivation of the one-component CPF
in the {\sl VdW\/} approximation.
Let us use it now for a {\em two-component\/} gas of spherical particles
with radii $R_1$ and $R_2$\,, respectively.
It is important to mention that each component may consist of several
particle species as long as these species have one common hard-core radius,
i.\,e.~the number of necessary {\sl VdW\/} components is determined by
the number of different excluded volume terms $b_{qq}$\,.
In the case of two radii the potential term (\ref{eq:pot1c}) becomes 
\begin{eqnarray}
  \label{eq:pot2c}
  \exp \l[ {\textstyle - \frac{U_{N_1+N_2}}{T} } \r]
  &=& {\textstyle \prod\limits_{i < j \le N_1} }
           \!\!\theta (\l| \vec x_{ij} \r| - 2 R_1) \times  \\ \nonumber
  & & \times {\textstyle \prod\limits_{k < \ell \le N_2} }
             \!\!\theta (\l| \vec x_{k \ell} \r| - 2 R_2) \times  \\
  & & \times {\textstyle \prod\limits_{\scriptstyle m \le N_1 \hfill \atop
                     \scriptstyle ~\, n \le N_2  \hfill} }
             \!\!\theta (\l| \vec x_{mn} \r| - (R_1+R_2)) ~.  \nonumber
\end{eqnarray}
The integration is carried out in the way described above;
e.\,g.~firstly over the coordinates of the particles of the second component,
then over those of the first component.
For the estimation of the excluded volume of a $k$-cluster now
{\em two different\/} particle sizes have to be considered.
One obtains
\begin{eqnarray}
  \lefteqn{Z(T,V,N_1,N_2)}  \nonumber \\
  &\quad \ge&
    \frac{\phi_1^{\,N_1}}{N_1!} \,\frac{\phi_2^{\,N_2}}{N_2!}
    \,\l\{{\textstyle \prod\limits_{i=0}^{N_1-1} }\!\l( V - 2b_{11}\,i \r) \r\}
    \times  \nonumber \\
  &  &
    {} \times \l\{{\textstyle \prod\limits_{j=0}^{N_2-1} }
    \! \l( V - 2b_{12}\,N_1 - 2b_{22}\,j \r) \r\}  \nonumber \\
  \label{eq:twoc_CPF}
  &\quad \cong&
    \frac{\phi_1^{\,N_1}}{N_1!}
    \,\frac{\phi_2^{\,N_2}}{N_2!} ~ V^{N_1+N_2} \times  \nonumber \\
  &  &
    {} \times \exp\l[{\textstyle  -\frac{N_1^{\,2} b_{11} + 2 N_1 N_2 b_{12}
                                         + N_2^{\,2} b_{22}}{V} }\r] ~,
\end{eqnarray}
where it is $\phi_q \equiv \phi(T;m_q,g_q)$\,, and
$2b_{pq} \equiv \frac{4 \pi}{3}\,(R_p+R_q)^3$ denotes the excluded volume
of a particle of the component $p$ seen by a particle of the component $q$
($p,~q=1, 2$ hereafter)\,.
Approximating the above exponent by $\exp[-x] \cong (1-x)$ 
yields the {\em simplest\/} possibility of a {\sl VdW\/} approximation
for the {\em two-component\/} CPF\,,
\begin{eqnarray}
  \label{eq:twoc_CPF_nl}
  \lefteqn{Z^{\rm\,nl}_{\rm VdW}(T,V,N_1,N_2)}  \nonumber \\
  &\quad \equiv&
    \frac{\phi_1^{\,N_1}}{N_1!}\,\frac{\phi_2^{\,N_2}}{N_2!} \times  \\
  &  &
    {}\times \l(V - \frac{N_1^{\,2}b_{11} +2N_1 N_2 b_{12} +N_2^{\,2}b_{22}}
                         {N_1+N_2}\r)^{\!N_1+N_2}  \nonumber \\
  \label{eq:twoc_CPF_nl-D}
  &\quad =&
    \frac{\phi_1^{\,N_1}}{N_1!} \,\frac{\phi_2^{\,N_2}}{N_2!} \times  \\
  &  &
    {}\times \l(V - N_1 b_{11} - N_2 b_{22}
                 + \frac{N_1\,N_2}{N_1+N_2}\,D\r)^{\!N_1+N_2} ~,
           \nonumber
\end{eqnarray}
where the non-negative coefficient $D$ is given by
\begin{equation}
  \label{eq:Def_D}
  D \equiv b_{11} + b_{22} - 2\,b_{12} ~.
\end{equation}
This approximation will be called the {\em non-linear\/} approximation
as the volume correction in (\ref{eq:twoc_CPF_nl-D}) contains non-linear
terms in $N_1, N_2$\,.
The corresponding pressure follows from the thermodynamical identity,
\begin{eqnarray}
  \label{eq:p-nl}
  \lefteqn{p^{\rm\,nl}(T,V,N_1,N_2)
           ~=~ p^{\rm\,nl}_1 + p^{\rm\,nl}_2} \\
  &\quad \equiv&
    \frac{T\,(N_1+N_2)}{V - N_1 b_{11} - N_2 b_{22}
                        + \frac{N_1\,N_2}{N_1+N_2}\,D}\, ~.  \nonumber
\end{eqnarray}
This canonical formula corresponds to the {\sl Lorentz-Berthelot\/} mixture
(without attraction terms) known from the theory of fluids \cite{Muen}\,.
It was postulated by {\sl van der Waals\/} \cite{VdW:1889} and studied
as well by {\sl Lorentz\/} \cite{Lor:1881} and
{\sl Berthelot\/} \cite{Berth:1898}\,.

The crucial step from the one- to the two-component gas is to include $b_{pq}$
terms ($p \neq q$) additionally to the $b_{qq}\equiv b|_{R=R_q}$ terms.
For the multi-component gas no further essential extension is necessary.
Consequently, the generalisation of the above procedure to the
multi-component case, i.\,e.~an arbitrary number of different
hard-core radii, is straightforward \cite{Zeeb:02}\,.

In Ref.~\cite{Gor:99} a {\em more involved\/} approximation has been
suggested for the two-component {\sl VdW\/} gas.
This follows from splitting the exponent in the CPF (\ref{eq:twoc_CPF}) by
introducing {\em two generalised\/} excluded volume terms $\tilde{b}_{12}$ and
$\tilde{b}_{21}$ (instead of a {\em single\/} and symmetric term $2\,b_{12}$)
for the mixed case,
\begin{eqnarray}
  \lefteqn{Z(T,V,N_1,N_2)}  \nonumber \\
  &\quad \cong&
    \frac{\phi_1^{\,N_1}}{N_1!} \,\frac{\phi_2^{\,N_2}}{N_2!}
    ~V^{N_1+N_2} \times  \\
  &  &
    {}\times \exp\l[{\textstyle -\frac{N_1^{\,2} b_{11}
                      + N_1 N_2 \l( \tilde{b}_{12} + \tilde{b}_{21} \r)
                      + N_2^{\,2} b_{22}}{V} }\r] ~,
    \nonumber
\end{eqnarray}
which leads to an alternative two-component {\sl VdW\/} CPF\,,
\begin{eqnarray}
\label{eq:twoc_CPF_lin}
    \lefteqn{Z_{\rm VdW}^{\rm\,lin}(T,V,N_1,N_2)}  \nonumber \\
  &\quad \equiv&
    \frac{\phi_1^{\,N_1}}{N_1!}
    ~ \l( V - N_1 b_{11} - N_2 \tilde{b}_{21} \r)^{N_1} \times  \\
  &  &
    {}\times \frac{\phi_2^{\,N_2}}{N_2!}
    ~ \l( V - N_2 b_{22} - N_1 \tilde{b}_{12} \r)^{N_2} ~.  \nonumber
\end{eqnarray}
Since the particle numbers $N_1, N_2$ appear solely linearly in the volume
corrections, these formulae will be referred to as the {\em linear\/}
approximation.
In this approximation one obtains \cite{Gor:99} for the pressure
\begin{eqnarray}
  \label{eq:p-lin}
  \lefteqn{p^{\rm\,lin}(T,V,N_1,N_2)
           ~=~ p_1^{\rm\,lin} + p_2^{\rm\,lin}}  \\
  &\quad \equiv&
    \frac{T \, N_1}{V - N_1 b_{11} - N_2 \tilde{b}_{21}}
    + \frac{T \, N_2}{V - N_2 b_{22} - N_1 \tilde{b}_{12}} \,~.  \nonumber
\end{eqnarray}

The choice of the generalised excluded volume terms $\tilde{b}_{pq}$ is
not unique in the sense that all choices which satisfy the basic
constraint $\tilde{b}_{12} + \tilde{b}_{21} = 2 b_{12}$ are consistent
with the second order virial expansion \cite{Gor:99}\,.
Therefore, additional conditions are necessary to fix these generalised
excluded volumes.
In Ref.~\cite{Gor:99} they were chosen as
\begin{eqnarray}
  \label{eq:bTi-s}
  \tilde{b}_{12} \equiv b_{11} \,\frac{2\,b_{12}}{b_{11}+b_{22}} ~,
  &\quad&
  \tilde{b}_{21} \equiv b_{22} \,\frac{2\,b_{12}}{b_{11}+b_{22}} ~.
\end{eqnarray}
For this choice, the linear approximation reproduces a traditional
{\sl VdW\/} gas behaviour, i.\,e.~{\em one-component-like\/}, in the two
limits $R_2=R_1$ and $R_2=0$ as readily checked.
The factor $2 b_{12}/(b_{11}+b_{22})=1-D/(b_{11}+b_{22})$ is always smaller
than unity for $R_1 \neq R_2$\,, consequently, the $\tilde b_{pq}$ terms
are smaller than the corresponding terms $b_{pp}$\,.
Note that there are many possible choices for $\tilde b_{12}$
and $\tilde b_{21}$\,, e.\,g.~additionally dependent on the particle numbers
$N_1$ and $N_2$\,, whereas the non-linear approximation (\ref{eq:twoc_CPF_nl})
contains no such additional parameters.

The formulae of the linear approximation are generally valid for {\em any\/}
choice of $\tilde{b}_{12}$ and $\tilde{b}_{21}$ satisfying the constraint
$\tilde{b}_{12}+\tilde{b}_{21}=2 b_{12}$\,.
In the following, however, we will restrict our study to the special choice
given in the Eqs.~(\ref{eq:bTi-s})\,.
The canonical (and grand canonical) formulae for the multi-component case
are given in Ref.~\cite{Gor:99}\,.

\subsection{Comparison of both \\ Two-component {\sl VdW\/} Approximations}

As the {\sl VdW\/} approximation is a low density approximation it is evident
that the linear and non-linear formulae are equivalent for such densities.
Deviations, however, occur at high densities, where any {\sl VdW\/}
approximation generally becomes inadequate.

The differences between both approximations result from the fact that the
linear pressure (\ref{eq:p-lin}) has two poles, $v^{\rm\,lin}_1=V$ and
$v^{\rm\,lin}_2=V$\,, whereas the non-linear pressure (\ref{eq:p-nl}) has
solely one pole, $v^{\rm\,nl}=V$\,.
For constant volume $V$ these poles define limiting densities,
e.\,g.~$\hat{n}_1=\max(N_1/V)$ as functions of $n_2=N_2/V$\,,
\begin{eqnarray}
  \label{eq:lim-n1,q-lin}
  v_q^{\rm\,lin}(N_1,N_2) = V & \quad \leadsto \quad &
    \hat{n}_1(n_2) \equiv \hat{n}_{1,q}^{\rm\,lin} (n_2)  \\
  \label{eq:lim-n1-nl}
  \mbox{or} \quad
  v^{\rm\,nl}(N_1,N_2) = V & \quad \leadsto \quad &
     \hat{n}_1(n_2) \equiv \hat{n}_1^{\rm\,nl} (n_2) ~,
\end{eqnarray}
which represent the domains of the two pressure formulae in
the $n_2$--$n_1$-plane.
The explicit fomulae are discussed in App.\ \ref{CE_appd}\,.

\begin{figure}[h]
  \begin{center}
    \psfig{file=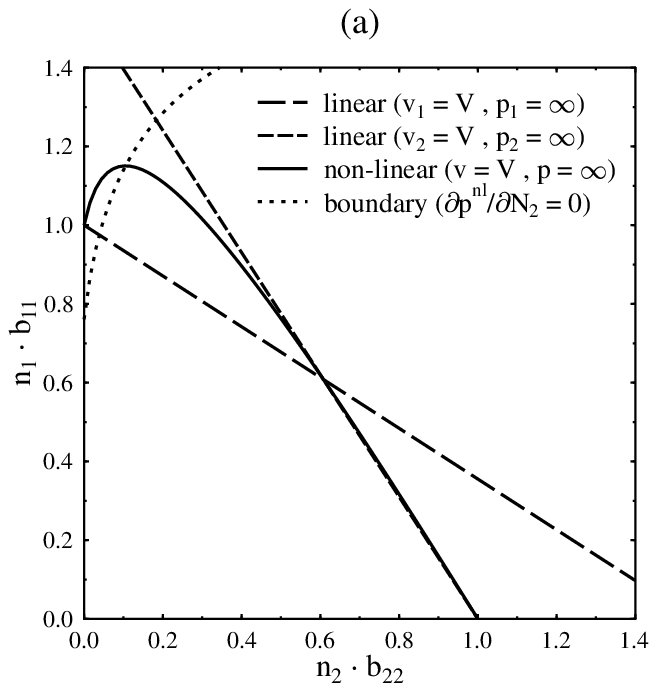, width=8.0cm}
    \psfig{file=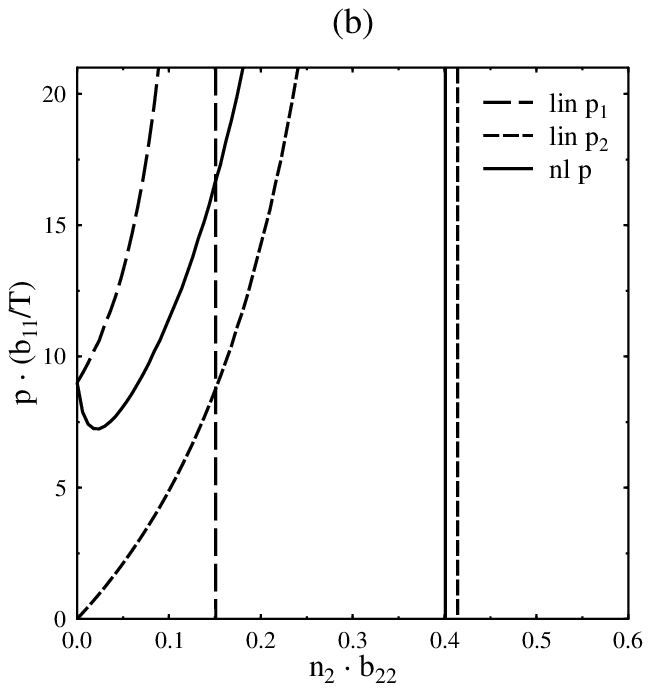, width=8.0cm}
  \end{center}
  \caption{
    \label{figs:1}
    (a) Domains of the linear and non-linear approximation
    for $R_2/R_1=0.4$\,:
    limiting densities $\hat{n}_1$ (isobars for $p_q(n_1,n_2)=\infty$)
    and the lower boundary $\hat{n}_1^{\rm\,nl,\,bd}$ to the region
    of negative partial derivatives of the non-linear pressure.
    The dashed lines correspond to the two poles of the linear pressure,
    and the solid line corresponds to the pole of the non-linear pressure.
    For given $n_2$ the possible densities $n_1^{\rm\,lin}$
    are below both dashed lines, whereas the possible densities
    $n_1^{\rm\,nl}$ are below the solid line.
    Negative derivatives $\pd p^{\rm\,nl}/\pd N_2 < 0$ occur only
    above the dotted line.
    \\
    (b) Pressure profiles in dimensionless units for $R_2/R_1 = 0.4$
    as in (a) at fixed $n_1 b_{11}= 0.9$\,.
    The dashed lines show the partial pressures of the linear approximation
    $p_1^{\rm\,lin}$ and $p_2^{\rm\,lin}$\,, while the solid line shows the
    total pressure of the non-linear approximation $p^{\rm\,nl}$
    with initial decrease due to negative $\pd p^{\rm\,nl} / \pd N_2$\,.
    }
\end{figure}

In Fig.~\ref{figs:1}\,(a) an example of these limiting densities is shown
for $R_2/R_1=0.4$\,.
It is clearly seen that the non-linear domain (below the solid line) is
larger than the linear domain (below both dashed lines), which is
generally the case for $R_2 \neq R_1$\,.
Especially for $R_2 \ll R_1$ the non-linear domain is distinctly larger
for high densities of the large component, $n_1 b_{11}>n_2 b_{22}$\,,
whereas both domains are similar for high densities of the small
component, $n_2 b_{22}>n_1 b_{11}$\,.

The linear approximation is constructed in traditional {\sl VdW\/} spirit;
the densities $n_q^{\rm\,lin}$ achieved in this approximation are
below the maximum density of the corresponding {\em one-component\/}
{\sl VdW\/} gas $\max(n_q^{\rm\,oc})=1/b_{qq}$\,, which is defined by
the pole of $p_q^{\rm\,oc}\equiv p_{\rm\,VdW}(T,V,N_q;b_{qq})$ from
Eq.~(\ref{eq:p_VdW})\,.

In the non-linear approximation, however, the possible densities of
the larger particles $n_1^{\rm\,nl}$ can exceed  $1/b_{11}$ due to the
occurence of negative partial derivatives of the pressure,
$\pd p^{\rm\,nl}/\pd N_2<0$\,.
In this context it is necessary to state that this behaviour does not
lead to a thermodynamical instability of the non-linear approximation
as proven in App.\ \ref{nl_appd}\,.
The linear approximation shows no such behaviour, it is always
$\pd p^{\rm\,lin}/\pd N_1>0$ and $\pd p^{\rm\,lin}/\pd N_2>0$\,.

The condition $\pd p^{\rm\,nl}/\pd N_2=0$ defines the boundary
$\hat{n}_1^{\rm\,nl,\,bd}(n_2)$ of the region of negative partial
derivatives of the non-linear pressure.
In Fig.~\ref{figs:1}\,(a) this boundary is shown by the dotted line
for $R_2/R_1=0.4$\,;
the values of $\pd p^{\rm\,nl}/\pd N_2$ are negative above this line.

Densities larger than $n_1^{\rm\,nl}=1/b_{11}$ can only occur, if $R_2$
is smaller than a critical radius,
\begin{equation}
  \label{eq:R2crit_CE}
  R_2 < R_{2,\rm\,crit}(R_1) = (\sqrt[3]{4} - 1)\,R_1 \approx R_1/1.7 ~.
\end{equation}
Then, the boundary $\hat{n}_1^{\rm\,nl,\,bd}(n_2)$ starts inside the
non-linear domain, see App.\ \ref{CE_appd} for details.

The reason for this behaviour is the ratio of the amounts of small
and large particles.
There are much more small than large particles in the system for
densities $n_1, n_2$ along the boundary $\hat{n}_1^{\rm\,nl,\,bd}(n_2)$
at high densities $n_1$\,: here, the fewer large particles are surrounded
by many small particles.
Therefore, the excluded volume interaction of the large particles
in the non-linear pressure (\ref{eq:p-nl}) is governed not by the
simple term $b_{11}$ but by the mixed term $b_{12}$\,, which is distinctly
smaller than $b_{11}$ for $R_2 \ll R_1$\,.
The maximum density achieved in the non-linear approximation
$\max(\hat{n}_1^{\rm\,nl})=4/b_{11}$ is obtained for $R_2 \to 0$ and 
$N_2 \gg N_1$\,, i.\,e.~these formulae go far {\em beyond\/} the
traditional {\sl VdW\/} results in the corresponding situation.

An example of pressure profiles for $p_1^{\rm\,lin}$\,, $p_2^{\rm\,lin}$ and
$p^{\rm\,nl}$ for $n_1 b_{11} = 0.9$ is shown in Fig.~\ref{figs:1}\,(b)\,,
where it is $R_2/R_1=0.4$ as in Fig.~\ref{figs:1}\,(a)\,.
The non-linear pressure (solid line) firstly decreases as the densities
$n_1, n_2$ correspond to the region of negative partial derivatives,
see Fig.~\ref{figs:1}\,(a)\,.
The partial pressures of the linear approximation are shown by dashed lines.
The non-linear domain is seen to be larger since it is one of the linear
partial pressures which diverges first for increasing $n_2$\,.

%

We conclude that the linear and non-linear approximation show a drastically
different behaviour for high values of the large component's density $n_1$\,.
In the linear approximation (\ref{eq:p-lin}) the possible density values
are below $1/b_{11}$ and $1/b_{22}$\,, respectively, and the derivatives
$\pd p^{\rm\,lin}/ \pd N_q$ are always positive.
Whereas in the non-linear approximation (\ref{eq:p-nl}) higher densities
$n_1 > 1/b_{11}$ are possible due to the occurence of negative derivatives
$\pd p^{\rm\,nl}/ \pd N_2 < 0$\,.
This may be considered as pathological -- or used as an advantage to
describe special situations, e.\,g.~densities
$1/b_{11}<n_1<\hat{n}_1^{\rm\,nl}$ for $R_2 \ll R_1$
(see App.~\ref{CE_appd}).

However, the use of any {\sl VdW\/} approximation is in principle problematic
for densities near $1/b_{qq}$\,.
For low densities the non-linear and linear approximation are
practically equivalent, and the non-linear approximation is preferable
since the formulae are essentially simpler.


\section{Grand Canonical Treatment}

Let us now turn to the grand canonical ensemble (GCE)\,.
The grand canonical partition function is built using the CPF\,,
\begin{eqnarray}
  \label{eq:GCPF}
  \lefteqn{{\cal Z}(T,V,\mu_1,\mu_2)}  \\
  &\quad =& {\textstyle \sum\limits_{N_1=0}^{\infty}
            \,\sum\limits_{N_2=0}^{\infty} }
            \exp\l[{\textstyle \frac{\mu_1 N_1 + \mu_2 N_2}{T} }\r]
            Z(T,V,N_1,N_2) ~,
            \nonumber
\end{eqnarray}
whereas the chemical potentials $\mu_1$ and $\mu_2$ correspond to the
components 1 and 2\,, respectively.
For the {\sl VdW\/} CPF (\ref{eq:twoc_CPF_nl-D}) or (\ref{eq:twoc_CPF_lin})
there are limiting particle numbers $\hat{N}_1(N_2)$ or $\hat{N}_2(N_1)$\,,
where each CPF becomes zero.
For this reason the above sum contains only a finite number of terms.
Then it can be shown that in the thermodynamical limit
(i.\,e.~in the limit $V\to\infty$ for $N_q/V={\rm const.}$)
the grand canonical pressure
$p(T,\mu_1,\mu_2)\equiv T\,\ln[{\cal Z}(T,V,\mu_1,\mu_2)]/V$
depends only on the {\em maximum term\/} of the double sum
(\ref{eq:GCPF})\,, where $N_1=\N_1$ and $N_2=\N_2$\,.
One obtains
\begin{eqnarray}
  \label{eq:p_GCE}
  \lefteqn{p(T,\mu_1,\mu_2)}  \\
  &\quad =& \lim_{V \to \infty} \frac{T}{V} \,
      \ln\biggl[\exp\l[{\textstyle \frac{\mu_1 \N_1 + \mu_2 \N_2}{T} }\r]
      Z(T,V,\N_1,\N_2) \biggr]~,  \nonumber
\end{eqnarray}
wheras $\N_1$ and $\N_2$ are the {\em average\/} particle numbers.

\subsection{The Two {\sl VdW} Approximations}

For the non-linear {\sl VdW\/} approximation (\ref{eq:twoc_CPF_nl-D})
the last expression takes the form
\begin{eqnarray}
  \label{eq:pnl_GCE}
  \lefteqn{p^{\rm\,nl}(T,\mu_1,\mu_2)} \\
  &\quad =&
    \lim_{V \to \infty} \frac{T}{V} \,
    \ln\l[{\textstyle \frac{A_1^{\,\N_1}}{\N_1!}
           \,\frac{A_2^{\,\N_2}}{\N_2!} }\r. \times  \nonumber \\
  &  &
    \l.{}\times \l({\textstyle  V - \N_1 b_{11} - \N_2 b_{22}
          + \frac{\N_1\,\N_2}{\N_1+\N_2}\,D }\r)^{\!\N_1+\N_2} \r]~,
     \nonumber
\end{eqnarray}
where $A_{q} = A(T,\mu_q;~m_q, g_q) \equiv \exp[\mu_q/T]\,\phi_q$\,.

The evaluation of both maximum conditions for the grand canonical pressure
\begin{eqnarray}
  0 &\stackrel{\textstyle !}{=}&
      \frac{\pd}{\pd \N_q}
      \l\{ \ln\l[{\textstyle  \frac{A_1^{\,\N_1}}{\N_1!}
                  \,\frac{A_2^{\,\N_2}}{\N_2!} }\r.\r. \times  \\
    &  &
      \l.\l.{}\times
      \l({\textstyle  V - \N_1 b_{11} - \N_2 b_{22}
          + \frac{\N_1 \, \N_2}{\N_1+\N_2}\,D }\r)^{\!\N_1+\N_2} \r] \,
      \r\} ~,  \nonumber
\end{eqnarray}
yields a system of two coupled transcendental equations,
\begin{eqnarray}
  \label{eq:xi1}
  \lefteqn{\xi_1^{\rm\,nl} (T, \mu_1 , \mu_2)}  \nonumber \\
  &\quad =&
    A_1 ~ \exp\l[{\textstyle  -(\xi_1^{\rm\,nl} + \xi_2^{\rm\,nl})\, b_{11}
                    + \frac{{\xi_2^{\rm\,nl}}^{\,2}}
                           {\xi_1^{\rm\,nl} + \xi_2^{\rm\,nl}}\,D }\r] \,,  \\
  \label{eq:xi2}
  \lefteqn{\xi_2^{\rm\,nl} (T, \mu_1 , \mu_2)}  \nonumber \\
  &\quad =&
    A_2 ~ \exp \l[{\textstyle  -(\xi_1^{\rm\,nl} + \xi_2^{\rm\,nl})\, b_{22}
                   + \frac{{\xi_1^{\rm\,nl}}^{\,2}}
                          {\xi_1^{\rm\,nl} + \xi_2^{\rm\,nl}}\,D }\r] \,,
\end{eqnarray}
where $\xi_1^{\rm\,nl}$ and $\xi_2^{\rm\,nl}$ are defined as
\begin{eqnarray}
  \label{eq:xi1_Def}
  \xi_1^{\rm\,nl}
    &\equiv& \frac{\N_1}{V - \N_1 b_{11} - \N_2 b_{22}
                         + \frac{\N_1 \, \N_2}{\N_1 + \N_2} \, D} \,~,  \\
  \label{eq:xi2_Def}
  \xi_2^{\rm\,nl}
    &\equiv& \frac{\N_2}{V - \N_1 b_{11} - \N_2 b_{22}
                         + \frac{\N_1 \, \N_2}{\N_1 + \N_2} \, D} \,~.
\end{eqnarray}
In the thermodynamical limit the average particle numbers $\N_1$ and $\N_2$
are proportional to $V$ as $\N_q = n_q^{\rm\,nl}\,V$\,.
Then the volume $V$ disappears in the definitions of $\xi_1^{\rm\,nl}$
and $\xi_2^{\rm\,nl}$ given by Eqs.~(\ref{eq:xi1_Def}, \ref{eq:xi2_Def})\,,
and they can be solved for either the density $n_1^{\rm\,nl}$ or
$n_2^{\rm\,nl}$\,,
\begin{eqnarray}
  \label{eq:n1}
  n_1^{\rm\,nl} (T, \mu_1 , \mu_2)
    &=& \frac{\xi_1^{\rm\,nl}}
             {1 + \xi_1^{\rm\,nl} b_{11} + \xi_2^{\rm\,nl} b_{22}
              - \frac{\xi_1^{\rm\,nl} \, \xi_2^{\rm\,nl}}
                     {\xi_1^{\rm\,nl} + \xi_2^{\rm\,nl}}\, D} \,~, \\
  \label{eq:n2}
  n_2^{\rm\,nl} (T, \mu_1 , \mu_2)
    &=& \frac{\xi_2^{\rm\,nl}}
             {1 + \xi_1^{\rm\,nl} b_{11} + \xi_2^{\rm\,nl} b_{22}
              - \frac{\xi_1^{\rm\,nl} \, \xi_2^{\rm\,nl}}
                     {\xi_1^{\rm\,nl} + \xi_2^{\rm\,nl}}\, D} \,~.
\end{eqnarray}
The $\xi_q^{\rm\,nl} = \xi_q^{\rm\,nl}(T,\mu_1,\mu_2)$ are the
solutions of the coupled Eqs.~(\ref{eq:xi1}) and (\ref{eq:xi2})\,,
respectively.

Hence, the pressure (\ref{eq:pnl_GCE}) can be rewritten in terms of
$\xi_1^{\rm\,nl}$ (\ref{eq:xi1}) and $\xi_2^{\rm\,nl}$ (\ref{eq:xi2})\,,
\begin{equation}
  \label{eq:pxi}
  p^{\rm\,nl} (T,\mu_1 ,\mu_2) = T\,\l(\xi_1^{\rm\,nl} + \xi_2^{\rm\,nl} \r) ~,
\end{equation}
supposed that Eqs.~(\ref{eq:n1}, \ref{eq:n2}) are taken into account.
If the definitions (\ref{eq:xi1_Def}) and (\ref{eq:xi2_Def}) are used
for $\xi_1^{\rm\,nl}$ and $\xi_2^{\rm\,nl}$\,,
the pressure formula (\ref{eq:pxi}) coincides with
the canonical expression (\ref{eq:p-nl}) for $N_1=\N_1$ and $N_2=\N_2$\,.

Since the formulation is thermodynamically self-consistent the
identity $n_q \equiv \pd p(T,\mu_1,\mu_2)/ \pd \mu_q$
leads to Eqs.~(\ref{eq:n1}, \ref{eq:n2}) as well.

The grand canonical formulae of the linear approximation
\cite{Gor:99} are obtained exactly as presented for the
non-linear case in Eqs.~(\ref{eq:pnl_GCE}--\ref{eq:pxi})\,.
In the linear case the two coupled transcendental equations are
\begin{eqnarray}
  \label{eq:xi1-lin}
  \xi_1^{\rm\,lin} (T, \mu_1 , \mu_2)
    &=& A_1 ~ \exp \l[ -\xi_1^{\rm\,lin}\, b_{11}
                       - \xi_2^{\rm\,lin}\, \tilde{b}_{12} \r] ~,  \\
  \label{eq:xi2-lin}
  \xi_2^{\rm\,lin} (T, \mu_1 , \mu_2)
    &=& A_2 ~ \exp \l[ -\xi_2^{\rm\,lin}\, b_{22}
                       - \xi_1^{\rm\,lin}\,  \tilde{b}_{21} \r] ~,
\end{eqnarray}
with the definitions
\begin{eqnarray}
  \label{eq:xi1-lin_Def}
  \xi_1^{\rm\,lin}
    &\equiv& \frac{\N_1}{V -\N_1 b_{11} -\N_2 \tilde{b}_{21}} \,~,  \\
  \label{eq:xi2-lin_Def}
  \xi_2^{\rm\,lin}
    &\equiv& \frac{\N_2}{V -\N_2 b_{22} -\N_1 \tilde{b}_{12}} \,~.
\end{eqnarray}
The expressions for the particle densities are found by solving
Eqs.~(\ref{eq:xi1-lin_Def}, \ref{eq:xi2-lin_Def}) for either
$n_1^{\rm\,lin}$ or $n_2^{\rm\,lin}$\,,
\begin{eqnarray}
  \label{eq:n1-lin}
  \lefteqn{n_1^{\rm\,lin} (T, \mu_1, \mu_2)}  \\  \nonumber
  &\quad =&
    \frac{\xi_1^{\rm\,lin} (1 + \xi_2^{\rm\,lin} \,[b_{22} - \tilde{b}_{21}]) }
         {1 + \xi_1^{\rm\,lin} b_{11} + \xi_2^{\rm\,lin} b_{22}
            + \xi_1^{\rm\,lin} \xi_2^{\rm\,lin}
              \,[b_{11} b_{22} - \tilde{b}_{12} \tilde{b}_{21}]} ~,  \\
  \label{eq:n2-lin}
  \lefteqn{n_2^{\rm\,lin} (T, \mu_1, \mu_2)} \\  \nonumber
  &\quad =&
    \frac{\xi_2^{\rm\,lin} (1 + \xi_1^{\rm\,lin} \,[b_{11} - \tilde{b}_{12}]) }
         {1 + \xi_1^{\rm\,lin} b_{11} + \xi_2^{\rm\,lin} b_{22}
            + \xi_1^{\rm\,lin} \xi_2^{\rm\,lin}
              \,[b_{11} b_{22} - \tilde{b}_{12} \tilde{b}_{21}]} ~.
\end{eqnarray}

For the linear approximation the pressure (\ref{eq:p_GCE}) can be rewritten
in terms of $\xi_1^{\rm\,lin}$ (\ref{eq:xi1-lin}) and
$\xi_2^{\rm\,lin}$ (\ref{eq:xi2-lin})\,,
\begin{eqnarray}
  \label{eq:pxi-lin}
  p^{\rm\,lin} (T, \mu_1 , \mu_2)
    = T \,\l(\xi_1^{\rm\,lin} + \xi_2^{\rm\,lin} \r) ~,
\end{eqnarray}
if Eqs.~(\ref{eq:n1-lin}, \ref{eq:n2-lin}) are taken into account,
like in the non-linear case.


\subsection{Comparison of both Approximations}

Let us brief\/ly return to the usual {\sl VdW\/} case,
the one-component case.
The corresponding transcendental equation is obtained from either
Eqs.~(\ref{eq:xi1}, \ref{eq:xi2}) or (\ref{eq:xi1-lin}, \ref{eq:xi2-lin})
by setting $R_1=R_2\equiv R$ and $A_1=A_2\equiv A$\,,
\begin{equation}
    \xi^{\rm\,oc}(T,\mu) = A \,\exp\l[ -\xi^{\rm\,oc}\,b \r] ~,
\end{equation}
whereas $b \equiv b_{11}=b_{22}$\,.
The {\em transcendental factor\/} $\exp[-\xi^{\rm\,oc}\,b]$ has the form of a
suppression term, and the solution $\xi^{\rm\,oc}\equiv p^{\rm\,oc}/T$
of this transcendental equation evidently decreases with increasing $b$
for constant $T$ and $\mu$\,.
Then in turn, the corresponding particle density
$n^{\rm\,oc}=\xi^{\rm\,oc}/(1+\xi^{\rm\,oc}\,b)$
is suppressed in comparison with the ideal gas due to the
lower $\xi^{\rm\,oc}$ {\em and\/} the additional denominator.
Thus, a suppressive transcendental factor corresponds to a suppression
of particle densities.

Now it can be seen from Eqs.~(\ref{eq:xi1}, \ref{eq:xi2}) and
(\ref{eq:xi1-lin}, \ref{eq:xi2-lin}) that the transcendental factors of
both {\em two-component\/} approximations contain as well this usual
{\em one-component-} or {\sl VdW}-{\em like\/} suppressive part
$\exp[-(p/T)\,b_{qq}]$\,.
But since it is $D \ge 0$ and $\tilde{b}_{pq} < b_{pp}$\,, respectively,
there is furthermore an attractive part in each corresponding transcendental
factor.

In the non-linear approximation the attractive part can even dominate the
suppressive part for the smaller component, e.\,g.~in Eq.~(\ref{eq:xi2})
for $R_2<R_1$\,.
Then the larger component can reach densities $n_1^{\rm\,nl}$ higher
than $1/b_{11}$\,, analogous to the CE\,.
A detailed discussion is given in App.\ \ref{GCE_appd}\,.

High densities in the canonical treatment correspond to large
values of the chemical potentials in the grand canonical treatment.
In the limit
\begin{equation}
  \label{eq:mu1lim}
  \mu_1/T \to \infty ~~ (T, \mu_2 = \mbox{const.})
    \quad \mbox{or} \quad
  \xi_1^{\rm\,nl} \to \infty
\end{equation}
the solution of Eq.~(\ref{eq:xi2})\,, $\xi_2^{\rm\,nl}$\,, can be enhanced
for increasing $\xi_1^{\rm\,nl}$ instead of being suppressed, if $R_2$ is
sufficiently small.
This may be called the {\em non-linear enhancement\/}.
The behaviour of the non-linear approximation in the limit (\ref{eq:mu1lim})
depends only on the ratio of the two radii $R_1/R_2$ and is characterised by
the coefficient
\begin{equation}
  \label{eq:a2-Def}
  a_2 \equiv \sqrt{D/b_{22}} - 1 ~.
\end{equation}
A negative $a_2$ relates to a suppressive transcendental factor in
Eq.~(\ref{eq:xi2})\,.
For equal radii $R_2=R_1$ it is $a_2=-1$\,, and the suppression is evidently
not reduced but {\sl VdW}-like.
For $-1<a_2<0$ this suppression is reduced, the most strongly for
$a_2\approx 0$\,.

In the case $a_2=0$ the suppression for $\xi_2^{\rm\,nl}$
(\ref{eq:xi2}) disappears in the limit (\ref{eq:mu1lim})\,,
on has $\xi_2^{\rm\,nl}\to A_2={\rm const.}$
This case provides the critical radius
$R_{2,\rm\,crit}$ (\ref{eq:R2crit_CE})\,.

For $a_2>0$ or $R_2<R_{2,\rm\,crit}$ the non-linear enhancement
of $\xi_2^{\rm\,nl}$ occurs for increasing $\xi_1^{\rm\,nl}$\,;
it is the stronger the larger $a_2$ is.
Then $n_1^{\rm\,nl}$ (\ref{eq:n1}) can exceed
$\max(n_1^{\rm\,oc})=1/b_{11}$\,,
whereas $n_2^{\rm\,nl}$ (\ref{eq:n2}) does not vanish
(see App.\ \ref{GCE_appd} for the explicit fomulae).
The density $\max(\hat{n}_1^{\rm\,nl})=4/b_{11}$ is achieved for
$a_2\to\infty$ or $R_2\to 0$\,.

The suppression in the transcendental factor of $\xi_1^{\rm\,nl}$
(\ref{eq:xi1}) is generally reduced for $R_2<R_1$\,, the more strongly
the smaller $R_2$ is, but there is no enhancement possible in the limit
(\ref{eq:mu1lim})\,.




\section{Relativistic Excluded Volumes}

In this section we will investigate the influence of re\-lativistic
effects on the excluded volumes of fast moving particles by accounting
for their ellipsoidal shape due to {\sl Lorentz\/} contraction.
In Ref.~\cite{Bug:99} a quite simple, ultra-relativistic approach has
been made to estimate these effects:
instead of ellipsoids two cylinders with the corresponding radii have been
used to calculate approximately the excluded volume term $b_{pq}$ for the
two-component mixture.
The resulting relativistic excluded volumes depend on the temperature
and contain the radii and the {\em masses\/} as parameters.
The simple, non-mixed term reads \cite{Bug:00}
\begin{eqnarray}
  \label{eq:bqqT}
  b_{qq} (T) &=& \alpha_{qq} \l(\frac{37 \pi}{9} \,\frac{\sigma_q}{\phi_q}
                   + \frac{\pi^2}{2} \r) R_q^{\,3} ~,
\end{eqnarray}
where $\sigma_q \equiv \sigma (T; m_q, g_q)$ denotes
the {\em ideal gas\/} scalar density,
\begin{equation}
  \sigma (T; m, g)
    = \frac{g}{2 \pi^2} \int\limits_0^{\infty} \dd k \, k^2 \,
      \frac{m}{E(k)} \,\exp\l[{\textstyle -\frac{E(k)}{T} }\r] ~.
\end{equation}

The expression for the mixed case can be derived similarly
from \cite{Bug:99}\,,
\begin{eqnarray}
  \label{eq:b12T}
  b_{12}(T) = \alpha_{12} \,
  & & \l\{{\textstyle
          \l(\frac{\sigma_1}{\phi_1}\,f_1
             + \frac{\pi^2}{4} \frac{R_2}{R_1}\r) R_1^{\,3} }\r. \\ \nonumber
  & & \l.{\textstyle
          {}+ \l(\frac{\sigma_2}{\phi_2}\,f_2
                 + \frac{\pi^2}{4} \frac{R_1}{R_2}\r) R_2^{\,3} }\r\} ~,
\end{eqnarray}
whereas the abbreviations $f_1$ and $f_2$ are dimensionless functions
of both radii,
\begin{eqnarray*}
  f_1 = {\textstyle \frac{\pi}{3} \!\l(2 + \frac{3R_2}{R_1}
                             + \frac{7R_2^{\,2}}{6R_1^{\,2}}\r) } ~,
  &\quad&
  f_2 = {\textstyle \frac{\pi}{3} \!\l(2 + \frac{3R_1}{R_2}
                             + \frac{7R_1^{\,2}}{6R_2^{\,2}}\r) }
 ~.
\end{eqnarray*}
The normalisation factors
\begin{eqnarray}
  \alpha_{11} = \alpha_{22}
    &=& {\textstyle \frac{16}{\textstyle \frac{37}{3} + \frac{3\pi}{2}} } ~,  \\
  \alpha_{12}
    &=& {\textstyle \frac{{\textstyle \frac{2\pi}{3} } \l(R_1+R_2\r)^3}
                         {\l(f_1 + {\textstyle \frac{\pi^2}{4}
                                     \frac{R_2}{R_1} }\r) R_1^{\,3}
                          + \l(f_2 + {\textstyle \frac{\pi^2}{4}
                                       \frac{R_1}{R_2} }\r) R_2^{\,3}} }
\end{eqnarray}
are introduced to normalise the ultra-relativistic approximations
(\ref{eq:bqqT}, \ref{eq:b12T}) for $T=0$ to the corresponding
non-relativistic results.
For the hadron gas, however, these {\sl Boltzmann\/} statistical formulae
will only be used at high temperatures, where effects of quantum statistics
are negligible.

Note that it is {\em not appropriate} to consider temperature dependent
hard-core radii $R_p(T)$ or $R_q(T)$ since the $b_{pq}(T)$ terms give
the {\sl Lorentz}-contracted excluded {\em volumes\/} and are involved
functions of $T, m_p, m_q, R_p$ and $R_q$\,.
However, for a given value of $b_{pq}(T)$ the necessary hard-core radii
$R_p$ and $R_q$ will obviously depend on the temperature.

It is evident that the formulae (\ref{eq:bqqT}, \ref{eq:b12T}) suffice
already for the multi-component case, because even a multi-component
{\sl VdW\/} formulation contains only $b_{pq}$ terms.

For both approximations the expressions for the pressure (\ref{eq:pxi})
or (\ref{eq:pxi-lin}) and corresponding particle densities
(\ref{eq:n1}, \ref{eq:n2}) or (\ref{eq:n1-lin}, \ref{eq:n2-lin})
remain unchanged.
However, due to the temperature dependence of the relativistic excluded
volumes the entropy density is modified
\begin{eqnarray}
  s(T,\mu_1,\mu_2)
    &\equiv& \frac{\pd}{\pd T} \, p(T,\mu_1,\mu_2)  \nonumber \\
  \label{eq:s}
    &\equiv& s_{\rm nrel}
             + s_{\rm rel} (\pd_T b_{11}, \pd_T b_{22}, \pd_T b_{12}) ~.
\end{eqnarray}
The additional term $s_{\rm rel}$ depends on temperature derivatives
of the relativistic excluded volumes,
$\pd_T b_{pq}\equiv\pd b_{pq}/\pd T$\,,
which represent their thermal compressibility.

Furthermore, the term $s_{\rm rel}$ generates additional terms
for the energy density, according to $e\equiv T s-p+\mu_1 n_1+\mu_2 n_2$\,.
In the non-linear approximation one obtains
\begin{eqnarray}
  \label{eq:e}
  \lefteqn{e^{\rm\,nl}(T,\mu_1,\mu_2)}  \nonumber \\
  &\quad=&
    n_1^{\rm\,nl} \, \frac{\epsilon_1}{\phi_1}
    + n_2^{\rm\,nl} \, \frac{\epsilon_2}{\phi_2}
    - (n_1^{\rm\,nl} + n_2^{\rm\,nl}) \ T^2 \times  \\  \nonumber
  &      &
    {}\times \l({\textstyle \xi_1^{\rm\,nl} \,\pd_T b_{11}
                + \xi_2^{\rm\,nl} \,\pd_T b_{22}
                - \frac{\xi_1^{\rm\,nl}\,\xi_2^{\rm\,nl}}
                       {\xi_1^{\rm\,nl} + \xi_2^{\rm\,nl}}\,\pd_T D }\r) ~,
\end{eqnarray}
and the linear approximation yields
\begin{eqnarray}
  \label{eq:e-lin}
  e^{\rm\,lin}(T,\mu_1,\mu_2)
    &=& n_1^{\rm\,lin} \, \frac{\epsilon_1}{\phi_1}
        + n_2^{\rm\,lin} \, \frac{\epsilon_2}{\phi_2}  \\
    && {}- n_1^{\rm\,lin} \, T^2 \, \l( \xi_1^{\rm\,lin} \, \pd_T b_{11}
             + \xi_2^{\rm\,lin} \, \pd_T \tilde{b}_{12} \r)  \nonumber \\
    && {}- n_2^{\rm\,lin} \, T^2 \, \l( \xi_2^{\rm\,lin} \,\pd_T b_{22}
             + \xi_1^{\rm\,lin} \,\pd_T \tilde{b}_{21} \r) ~,  \nonumber
\end{eqnarray}
whereas $\epsilon_q\equiv\epsilon(T;m_q,g_q)$ denotes the {\em ideal gas\/}
energy density
\begin{equation}
  \epsilon (T; m, g) = \frac{g}{2 \pi^2} \int\limits_0^{\infty} \dd k\,k^2\,
    E(k)\,\exp \l[{\textstyle -\frac{E(k)}{T} }\r] ~.
\end{equation}
The additional terms in the entropy density (\ref{eq:s}) and in the
energy density (\ref{eq:e}) or (\ref{eq:e-lin}) which contain temperature
derivatives do evidently not occur in the case of the usual non-relativistic,
i.\,e.~constant excluded volumes.

Let us now study the hadronic equation of state generated
by each of the two-component {\sl VdW\/} approximations and 
their modifications due to relativistic excluded volumes.
When used to describe {\em hadronic\/} particles, the hard-core radii $R_q$
should be considered as {\em parameters\/} rather than particle radii.
We identify the first component as nucleons ($m_1 \equiv m_{\rm n}=939$ MeV\,,
$\mu_1 \equiv \mu_{\rm n} = \mu_{\rm B}$ and $g_1 \equiv g_{\rm n}=4$
for symmetric nuclear matter) and the second as pions
($m_2 \equiv \overline{m}_\pi=138$ MeV\,, $\mu_2 \equiv \mu_\pi = 0$ and
$g_2 \equiv g_\pi=3$)\,.
Quantum statistical effects other than the degeneracy factors $g_q$ are
neglected. 
To reproduce experimental data, however, it would be necessary to consider
all hadrons and hadronic resonances as well as the contributions from
hadronic decays into daughter hadrons.

\begin{figure}[h]
  \begin{center}
    \psfig{file=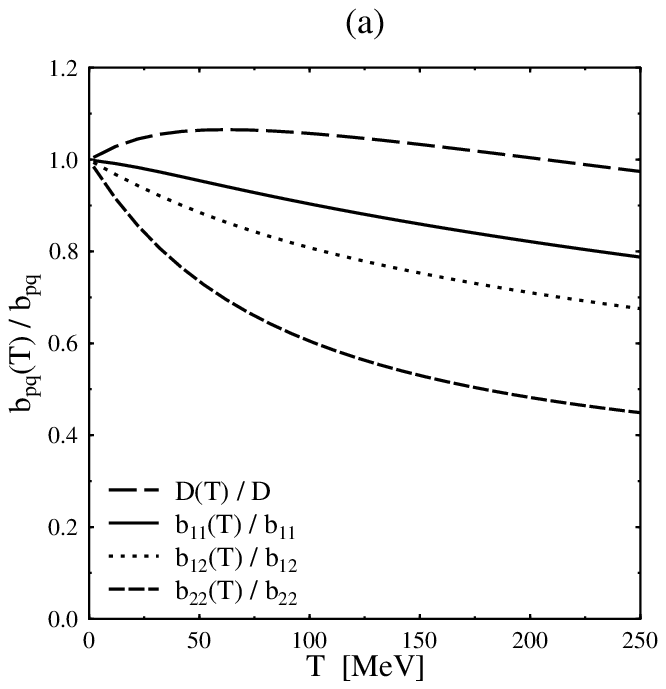, width=8.0cm}
    \psfig{file=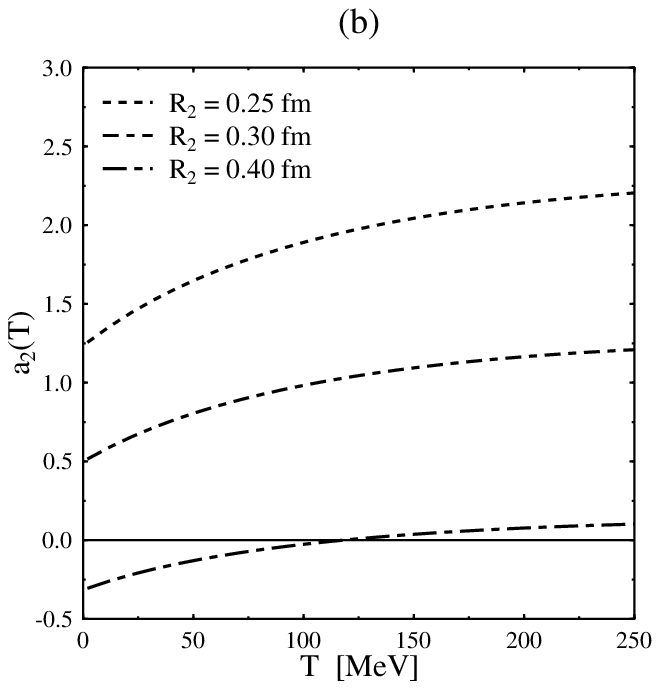, width=8.0cm}
  \end{center}
  \caption{
    \label{figs:2}
    Temperature dependence of the relativistic excluded volume terms
    for $m_1=m_{\rm n}, m_2=\overline{m}_\pi, R_1=0.6$ fm\,.  \\
    (a) Relative values for $R_2=0.3$ fm\,:
        $b_{11}(T)/b_{11}$\,, $b_{12}(T)/b_{12}$\,,
        $b_{22}(T)/b_{22}$ and $D(T)/D$ (solid line, dotted line,
        short and long dashes, respectively).
        The relativistic excluded volume of light species ($b_{22}(T)/b_{22}$)
        is affected more strongly by temperature. \\
    (b) The characteristic coefficient of the non-linear approximation
        $a_2(T)=(\sqrt{D(T)/b_{22}(T)}-1)$ for various
        $R_2=0.25$\,, 0.3 and 0.4 fm\,.
        The non-linear enhancement ($a_2(T)>0$) becomes stronger due to
        the decrease of the relativistic excluded volumes with increasing
        temperature.
    }
\end{figure}

For some examples the temperature dependence of the relativistic excluded
volumes is shown in Fig.~\ref{figs:2}\,(a)\,, given in units of the
corresponding non-relativistic terms, $b_{pq}=b_{pq}(0)$\,.
The solid line and the short dashes show the basic excluded volumes
$b_{11}(T)$ and $b_{22}(T)$\,, respectively.
In these relative units the decreases of $b_{11}(T)$ and $b_{22}(T)$
depend only on the corresponding masses.
It is apparent that the pion excluded volume $b_{22}(T)$ is affected much
stronger than the excluded volume of the nucleons, $b_{11}(T)$\,.
The dotted line shows the mixed volume term $b_{12}(T)$\,,
and the long dashes show the compound volume term
$D(T) \equiv b_{11}(T)+b_{22}(T)-2b_{12}(T)$\,.
These two terms depend obviously on both masses and both radii.


The curves for the generalised excluded volume terms of the linear
approximation $\tilde{b}_{12}(T)$ and $\tilde{b}_{21}(T)$ behave
similarly to $b_{12}(T)$\,.

Introducing the relativistic excluded volumes $b_{pq}(T)$\,, however,
has two effects.
First, the maximum densities become larger since it is generally
$1/b_{qq}(T)>1/b_{qq}$ as seen in Fig.~\ref{figs:2}\,(a)\,.
Furthermore, the balance between the lighter and the heavier species is
changed because the lighter species is affected more than the heavier
at the same temperature:
For the above parameters it is $b_{22}(T)/b_{22}\le b_{11}(T)/b_{11}$\,.

\newpage
In the non-linear approximation this balance is characterised by the
coefficient $a_2$ defined by Eq.~(\ref{eq:a2-Def})\,.
In Fig.~\ref{figs:2}\,(b) the temperature dependence of
$a_2(T)\equiv(\sqrt{D(T)/b_{22}(T)}-1)$ is shown for three different
values of $R_2$\,.
The {\em relativistic\/} coefficient $a_2(T)$ increases with $T$\,,
i.\,e.~the non-linear enhancement becomes stronger for higher temperatures.
For some values of $R_2$\,, e.\,g.~$R_2=0.4$ fm\,, a primary suppression
$a_2(0)\equiv a_2<0$\,, turns into an enhancement $a_2(T)>0$\,, when the
temperature is sufficiently high.
For temperature dependent excluded volumes $R_{2,\rm\,crit}$ looses its
meaning; here, only $a_2(T)>0$ is the valid condition for the
occurrence of the non-linear enhancement or densities
$n_1^{\rm\,nl}>1/b_{11}(T)$\,.

The linear coefficient, $\tilde{a}_2(T)=-2b_{12}(T)/(b_{11}(T)+b_{22}(T))$\,,
is not strongly affected by temperature for the above choice of hadronic
parameters:
It increases slightly with $T$ but remains negative.
Hence, changes in the balance between the lighter and the heavier species
play a minor role for the linear approximation.

\begin{figure}[h]
  \begin{center}
    \psfig{file=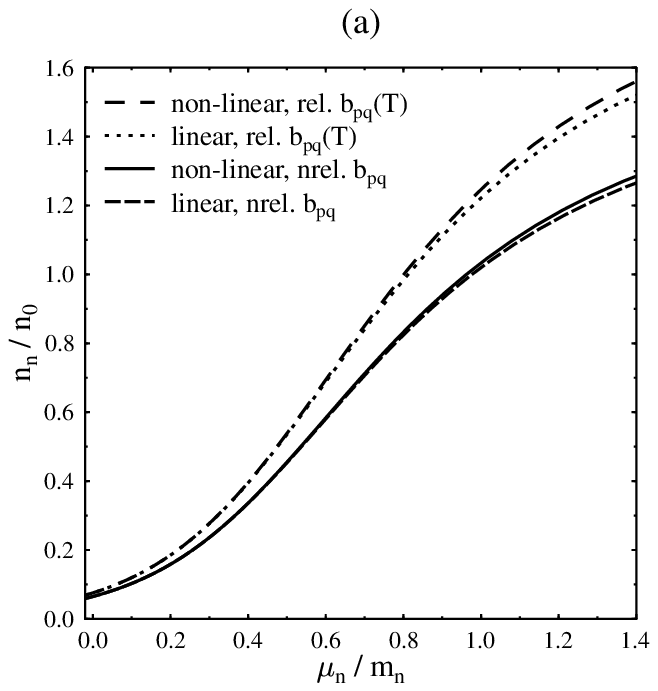, width=8.0cm}
    \psfig{file=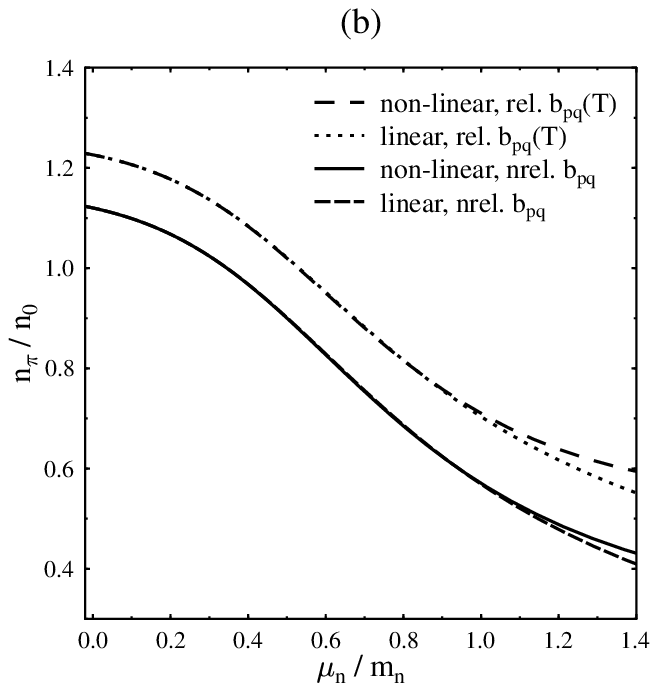, width=8.0cm}
  \end{center}
  \caption{
    \label{figs:3}
    Comparison of the model predictions for the nucleon (a) and
    pion (b) density $n_{\rm n}$ and $n_\pi$\,, respectively,
    vs.~$\mu_{\rm n}/m_{\rm n}$
    ($R_1=0.6$ fm\,, $R_2=0.3$ fm and $T=185$ MeV\,,
    densities in units of $n_0=0.16$ fm$^{-3}$).
    In both figures the two upper lines correspond to relativistic
    excluded volumes $b_{pq}(T)$ and the two lower lines to non-relativistic
    excluded volumes $b_{pq}$\,.
    The results of the linear and non-linear approximation coincide --
    only for extremely large $\mu_{\rm n}/m_{\rm n}$ the non-linear results
    lie slightly higher than the corresponding linear results.
    The deviations due to relativistic excluded volumes are significant.
    }
\end{figure}

Particle densities for nucleons and pions in units of
$n_0=0.16$ fm$^{-3}$ vs.~$\mu_1/m_1 \equiv \mu_{\rm n}/m_{\rm n}$ are
shown in Figs.~\ref{figs:3}\,(a) and (b) for $T = 185$ MeV\,.
The linear and non-linear results are shown for constant excluded volumes
with short dashes and solid lines, respectively, and further for
relativistic excluded volumes with dotted lines and long dashes,
respectively.
At this high temperature the relativistic results are significantly higher
than the non-relativistic result.
A difference between the linear and the non-linear approximation due to
the non-linear enhancement becomes noticeable only for high
$\mu_{\rm n}/m_{\rm n} > 0.8$\,.
Thus, for $R_{\rm n}=R_1=0.6$ fm from above, the linear and non-linear
approximation are practically equivalent for nucleon densities below
$n_{\rm n} \approx 0.8\,n_0$\,, i.\,e.~for densities below about $n_1 \approx 1/(2\,b_{11})$\,.
On the other hand, due to the strong decrease of $b_{22}(T)$ with
increasing temperature, the influence of the relativistic excluded volumes
is essential for temperatures of the order of $T \approx m_\pi$\,.

The presence of the additional terms containing temperature
derivatives in the energy density (\ref{eq:e}) or (\ref{eq:e-lin})
makes it impossible to convert a {\sl VdW\/} gas with relativistic
excluded volumes into a gas of free streaming particles.
Therefore, it is problematic to use these formulae for the post-freeze-out
stage.
For the latter the quantities of the free streaming particles without any
interaction should be used, see discussion in \cite{Bugs,Mag_Andl}
and references therein.
However, these equations of state may be used to describe the stage between
chemical and thermal freeze-out, i.\,e.~a pre-freeze-out stage in terms of
Refs.~\cite{Bugs,Mag_Andl}\,.
This is examplified in the next section.

\newpage

\section{Hard-core Radii \\ from Particle Yield Ratios}

As a simple application of the equations of state presented above,
let us re-evaluate the thermal model fit parameters for particle yield
ratios of Ref.~\cite{Yen:97}\,, namely the hard-core radii of pions $R_\pi$
and other hadrons $R_{\rm o}$\,.
A two-component {\sl VdW\/} excluded volume model has been used there
to explain the pion abundance in A+A-collisions by a smaller hard-core
radius for the pions than for the other hadrons.
The ratios has been fitted to BNL AGS (Au+Au at 11 A\,GeV) and CERN SPS
(Pb+Pb at 160 A\,GeV) data \cite{PrQM:96} within a thermal model, including
all resonances up to 2 GeV and using quantum statistics.

The applied model, however, corresponds to the incorrect separated model
as pointed out in Sect.~I\,.
For convenience we give these formulae 
in {\sl Boltzmann\/} approximation.
Within the previously defined  notation the two coupled transcendental
equations read
\begin{eqnarray}
  \label{eq:xi1-sp}
  \xi_1^{\rm\,sp} (T, \mu_1, \mu_2)
    &=& A_1 \,\exp \l[ -(\xi_1^{\rm\,sp} + \xi_2^{\rm\,sp})\, b_{11}\r] ~,  \\
  \label{eq:xi2-sp}
  \xi_2^{\rm\,sp} (T, \mu_1, \mu_2)
    &=& A_2 \,\exp \l[ -(\xi_1^{\rm\,sp} + \xi_2^{\rm\,sp})\, b_{22} \r] ~,
\end{eqnarray}
wheras $p^{\rm\,sp}(T,\mu_1,\mu_2)=T\,(\xi_1^{\rm\,sp}+\xi_2^{\rm\,sp})$\,.
In this context $A_1$ represents a sum over the contributions of all hadron
species but pions, while $A_2$ corresponds to the pions only.

The expressions for the particle densities are obtained from
$n_q^{\rm\,sp} \equiv \pd p^{\rm\,sp}/ \pd \mu_q$\,,
\begin{eqnarray}
  n_1^{\rm\,sp} (T, \mu_1, \mu_2)
    &=& \frac{\xi_1^{\rm\,sp}}
             {1 + \xi_1^{\rm\,sp} b_{11} + \xi_2^{\rm\,sp} b_{22}}\, ~,  \\
  \label{eq:n2-sp}
  n_2^{\rm\,sp} (T, \mu_1, \mu_2)
    &=& \frac{\xi_2^{\rm\,sp}}
             {1 + \xi_1^{\rm\,sp} b_{11} + \xi_2^{\rm\,sp} b_{22}}\, ~.
\end{eqnarray}
Solving these equations for $\xi_1^{\rm\,sp}$ and $\xi_2^{\rm\,sp}$
one recovers the canonical pressure formula of the separated model
(\ref{eq:p-sp_CE}) as announced in Sect.~II\,.

Due to the separation of both components in this model there is no excluded
volume term $b_{12}$ for the interaction between different components at all.
This is an essential difference to both the linear and the non-linear
approximation.
Note that the separated model is {\em not\/} a two-component
{\sl VdW\/} {\em approximation\/} because it cannot be obtained
by approximating the CPF (\ref{eq:twoc_CPF})\,.

The transcendental factors of the formulae (\ref{eq:xi1-sp}, \ref{eq:xi2-sp})
exhibit a constant {\sl VdW}-like suppression $\exp[-(p/T)\,b_{qq}]$\,.
There is a reduction of this suppression in the linear and in the
non-linear approximation, as discussed in Sect.~III\,.
The {\sl VdW}-like suppression is reduced, if $b_{12}$ appears in the
corresponding formulae since $b_{12}$ is smaller than $b_{11}$ for $R_2<R_1$\,.
It is evident that the deviation of the linear and non-linear approximation
from the separated model is the larger the more $R_1$ and $R_2$ differ from
each other.

In the first step of the fit procedure of Ref.~\cite{Yen:97}
only the hadron ratios excluding pions have been taken to find the
freeze-out parameters.
For AGS  $T \approx 140$ MeV\,, $\mu_{\rm B} \approx 590$ MeV and for SPS
$T \approx 185$ MeV\,, $\mu_{\rm B} \approx 270$ MeV have been obtained.
In the second step, a parameter introduced as the
{\em pion effective chemical potential\/} $\mu_\pi^{\,*}$ has been fitted to
the pion-to-hadron ratios.
Using {\sl Boltzmann\/} statistics it can be shown that the pion enhancement
is thoroughly regulated by the value of $\mu_\pi^{\,*}$ \cite{Yen:97}\,;
one has obtained $\mu_\pi^{\,*} \approx 100$ MeV for AGS and
$\mu_\pi^{\,*} \approx 180$ MeV for SPS data, respectively.

The pion effective chemical potential depends explicitly on the
excluded volumes but also on the pressure.
The pressure itself is a transcendental function depending solely on the
excluded volumes since $T$ and $\mu_{\rm B}$ are already fixed by step one.
In Ref.~\cite{Yen:97} the formula 
$\mu_\pi^{\,*} \equiv (v_{\rm o} - v_\pi)\,p(v_{\rm o}, v_\pi)$
has been obtained for the separated model, where $v_\pi \equiv b_{22}$
and $v_{\rm o} \equiv b_{11}$ are the excluded volumes corresponding
to the hard-core radii of pions $R_\pi \equiv R_2$ and other hadrons
$R_{\rm o} \equiv R_1$\,, respectively.
Thus, the $\mu_\pi^{\,*}$ values for AGS and SPS data define two curves
in the $R_\pi$--$R_{\rm o}$-plane.
The main conclusion of Ref.~\cite{Yen:97} is that the intersection point
of these two curves ($R_\pi=0.62$ fm\,, $R_{\rm o}=0.8$ fm) gives the correct
pair of hard-core radii for pions and for the other hadrons, i.\,e.~AGS and
SPS data are fitted simultaneously within the 
applied model.

In Ref.~\cite{BrM:99} these values of $R_{\rm o}$ and $R_\pi$ have been
criticised for being unreasonably large.
There, a complete fit of solely SPS data has been performed within
a separated model.
The best fit has been obtained for equal hard-core radii,
$R_\pi=R_{\rm o}=0.3$ fm\,, motivated by nucleon scattering data.
Good agreement has been found as well for a {\em baryon\/} hard-core radius,
$R_{\rm Bar}=0.3$ fm\,, and a common hard-core radius for {\em all mesons\/},
$R_{\rm Mes}=0.25$ fm\,, choosen in accord with the above ratio
of radii, $R_{\rm o}/R_\pi=0.8/0.62$\,.
Larger hard-core radii, especially those of Ref.~\cite{Yen:97}\,, are
quoted as giving distinctly worse agreement.

\begin{figure}[h]
  \begin{center}
    \psfig{file=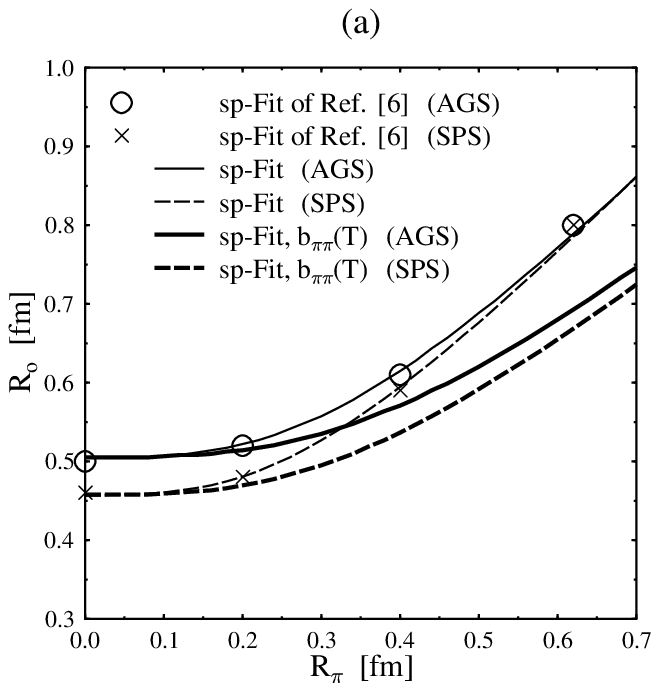, width=8.0cm}
    \psfig{file=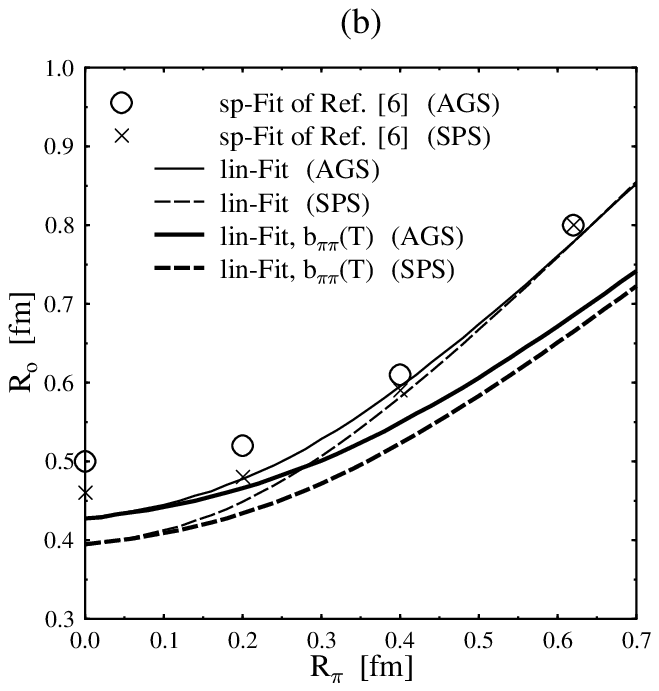, width=8.0cm}
  \end{center}
  \caption{
    \label{figs:4}
    Fits of particle yield ratios for AGS and SPS
    data \protect\cite{PrQM:96} with the separated model
    and the linear approximation.
    The thin lines show the fits for the separated model (a)
    and the linear approximation (b)\,;
    the non-linear approximation gives identical results for
    the latter case.  \\
    The thick lines in (a) and (b) show the corresponding curves for
    relativistic excluded volumes $v_\pi=b_{22}(T)$\,:
    there is no intersection for either of the three models.
    In both figures the results of the fit from Ref.~\protect\cite{Yen:97}
    for AGS and SPS data are indicated by circles and crosses, respectively.
    }
\end{figure}

Assuming the validity of {\sl Boltzmann\/} statistics,
we have re-calculated the $R_{\rm o}(R_\pi)$-curves for the above
$\mu_\pi^{\,*}$ values;
firstly in the separated model (\ref{eq:xi1-sp}--\ref{eq:n2-sp})\,,
i.\,e.~as presented in \cite{Yen:97}\,.
The resulting curves, shown as thin lines in Fig.~\ref{figs:4}\,(a)\,,
naturally match the results of the underlying fit of Ref.~\cite{Yen:97}\,,
which are indicated by markers.

Then we have considered the linear and the non-linear approximation.
Due to the occurence of $b_{12}$ terms in these two cases, both functional
forms of $\mu_\pi^{\,*}$ are essentially different from the separated case.
Consequently, the shapes of the $R_{\rm o}(R_\pi)$-curves are different
as well.
We find distinct deviations from the separated model, especially for
$R_\pi \to 0$\,, and the values for the intersection point are slightly
lower; see thin lines in Fig.~\ref{figs:4}\,(b) for the linear exrapolation.
The  non-linear approximation gives identical results for this purpose because
the hadron densities are too small for a noticeable non-linear enhancement.


The crucial point is now to turn on the relativistic temperature
dependence of the excluded volumes.
To keep the analysis simple we treat only pions this way since they
give the strongest effect.

\newpage
Although the other hadrons are assumed to have equal hard-core radii,
their relativistic excluded volumes would be different for $T > 0$\,,
according to their different masses.
To check the influence of relativistic excluded volumes for {\em all\/}
particles, we have used {\em one average\/} mass of 1~GeV for
{\em all other\/} hadrons.
The corresponding change in the $R_{\rm o}$-values are below 5\%\,.

The results of the fit for relativistic excluded volumes for pions
are shown in Figs.~\ref{figs:4}\,(a) and (b) as thick lines.
Though this approach is more realistic, there is {\em no\/} intersection
point for {\em any\/} of the three models even for very large radii
$R_{\rm o}, R_\pi \gg 0.5$ fm\,.
For the approximated case of a single averaged hadron mass there is no
intersection either.
Because of the different freeze-out temperatures for AGS and SPS
the $v_\pi(T) \equiv b_{22}(T)$ values are changed differently
in both cases, and so are the scales for the corresponding $R_\pi$\,.

Due to the errors in experimental data one ought to obtain a corridor instead
of a curve for each set of data.
Consequently, the particle yield ratios can be reproduced well by
e.\,g.~$R_{\rm o} \approx 0.4$ fm\,, $R_\pi \approx 0.2$ fm or larger values
for any of the models with relativistic excluded volumes.
Therefore, we conclude that the fit procedure proposed in Ref.~\cite{Yen:97}
is {\em not\/} suitable to find a {\em unique\/} pair of hard-core radii for
pions and other hadrons, as long as a best fit is searched for just two sets
of data of particle yield ratios.
The use of a relativistic excluded volume for pions along with a correct
approximation reduce the value of the necessary nucleon hard-core radius
essentially towards more realistic values.


\section{Summary}

In the present work 
several equations of state for the two-component {\sl Van der Waals\/}
excluded volume model are derived and investigated.
We have discussed two essentially different formulations, the linear and
the non-linear approximation.

The non-linear approximation is the simplest possibility.
Here, the large component can reach higher densities $n_1$ than the
usual limiting {\sl VdW\/} density $1/b_{11}$\,, if the other component
has a suffiently small hard-core radius, $R_2 < R_{2,\rm\,crit}$\,.
%
In the linear approximation the densities cannot exceed the usual
limiting {\sl VdW\/} densities $1/b_{11}$ and $1/b_{22}$\,, but
generalised excluded volume terms have to be introduced.

For both approximations the suppression factors of the grand canonical
formulae contain a {\sl VdW}-like term, proportional to
$\exp[-(p/T)\,b_{qq}]$\,, which however is reduced non-trivially.
In the linear case there is a slight reduction, wheras in the non-linear case
this reduction can turn the suppression even into an enhancement for the
smaller component, which leads to exceeding of $1/b_{11}$ for the
density of the larger component $n_1$\,.

The commonly used formulae of the separated model are shown to be not
suitable for the two-component case, because they correspond to a system
where both components are separated from each other and cannot mix.
In this model the grand canonical suppression factor is just {\sl VdW}-like
and has no reduction of the suppression.

Furthermore relativistic, i.\,e.~{\sl Lorentz}-contracted, excluded volumes
have been introduced.
Naturally, the relativistic excluded volume per particle decreases with
rising temperature.
This effect is the stronger the lighter the particle species is.
The suppression of particle densities in {\sl VdW\/} models is lower for
a component of smaller excluded volume in comparison with a component
of larger excluded volume.
Therefore, the temperature dependence of the relativistic excluded volumes
causes a reduction of the particle densities suppression.

The full equations of state have been presented, for both the linear
and non-linear approximation, with constant and with relativistic
excluded volumes.
For the entropy density and energy density there are additional terms
containing temperature derivatives of the relativistic excluded volume
terms due to their 'thermal compressibility'.
In comparison with the non-relativistic case, the expressions for the
pressure and the particle densities remain unchanged, but the possible
range of values is obviously wider, since it is generally
$1/b_{11}(T) \ge 1/b_{11}$ and $1/b_{22}(T) \ge 1/b_{22}$\,.

As an application of the derived formulations a fit of particle yield
ratios for SPS and AGS has been re-evaluated.
In Ref.~\cite{Yen:97} this fit had been done in a separated model
by adjusting the hard-core radii for the pions $R_\pi$ and for the other
hadrons $R_{\rm o}$\,.
The results of the new fit are essentially different from the separated
model but coincide for both the linear and non-linear approximation.
The picture changes drastically, however, if relativistic excluded volumes
are adopted for pions.
The basic idea of the fit -- one pair of hard-core radii suffices to fit
AGS and SPS data simultaneously -- does not lead to a result anymore.
This is the case for the separated model and for both approximations,
linear and non-linear.
Experimental uncertainties lead to a {\em region\/} of possible values
in the $R_{\rm o}$--$R_\pi$-plane;
one could describe the data for $R_{\rm o} \ge 0.4$ fm and
$R_\pi \ge 0.2$ fm\,.

We conclude that there are two causes of an {\em enhancement\/} of particle
densities, e.\,g.~the thermal pion abundance, in {\sl VdW\/} descriptions:
First, the density suppression is generally lower for the smaller component
in two-component models.
Second, there is a further reduction of the density suppression
due to the relativistic excluded volumes.
The latter are essentially smaller for light hadron species than for
heavy species, especially for temperatures $T \gg 50$ MeV\,.

When applied to the hadron gas, the linear and non-linear results almost
coincide for nucleon densities up to $n_0 \approx 0.16$ fm$^{-3}$ (for
$R_{\rm o} \le 0.6$ fm) since the non-linear enhancement does not appear
there, but the deviation from the incorrect separated model is distinct.
However, the formulae of the non-linear approximation are essentially
simpler than these of the linear approximation.

The influence of relativistic effects on the excluded volumes becomes
indispensable for temperatures typical for heavy ion collisions.
Therefore, it is necessary to include a correct two- or multi-component
{\sl VdW\/} approximation -- linear or non-linear -- as well as
relativistic excluded volumes in future calculations.


\acknowledgements

The authors thank M.\,I.~Gorenstein and D.\,H. Rischke for useful discussions
and St.~Hofmann for valuable comments.

The financial support of the NATO Linkage Grant PST.CLG.976950 is
acknowledged.


\appendix

\section{The Two VdW Approximations in the CE}
\label{CE_appd}

In the following we will study the differences between the linear and
the non-linear approximation:
the total excluded volumes $v_q=v_q(N_1,N_2)$ of the corresponding
partial pressures.
In the linear pressure formula (\ref{eq:p-lin}) each component has its own
total excluded volume given by
\begin{equation}
  \label{eq:v-lin1}
  v^{\rm\,lin}_1 \equiv N_1 b_{11} + N_2 \tilde{b}_{21} ~, \quad
  v^{\rm\,lin}_2 \equiv N_1 \tilde{b}_{12} + N_2 b_{22} ~,
\end{equation}
whereas in the non-linear pressure formula (\ref{eq:p-nl}) there is
a {\em common\/} total excluded volume for both components
\begin{equation}
  \label{eq:v-nl}
  v^{\rm\,nl}_1 = v^{\rm\,nl}_2 = v^{\rm\,nl}
    \equiv N_1 b_{11} + N_2 b_{22}
           - {\textstyle \frac{N_1\,N_2}{N_1+N_2} } D ~.
\end{equation}
It can be readily checked that it is either
$v^{\rm\,lin}_1 \le v^{\rm\,nl} \le v^{\rm\,lin}_2$ or 
$v^{\rm\,lin}_1 \ge v^{\rm\,nl} \ge v^{\rm\,lin}_2$\,,
i.\,e.~the pole of the non-linear pressure 
always lies between both poles of the linear pressure. 
Hence, there are values $N_1, N_2$\,, where the non-linear pressure
is still finite, but the linear pressure formula is yet invalid since
one of the partial pressures has already become infinite.
Consequently, the domain of the non-linear approximation is larger.

For given $V$ the two domains can be expressed by the limiting densities
(\ref{eq:lim-n1,q-lin}) and (\ref{eq:lim-n1-nl})\,, which are defined
by the poles $v_q(N_1,N_2)=V$ of the corresponding pressure.
In the linear approximation one obtains the expressions
\begin{eqnarray}
  \hat{n}_{1,1}^{\rm\,lin} (n_2)
    = \frac{1 - \tilde{b}_{21} n_2}{b_{11}}\, ~,
  &\quad&
    \hat{n}_{1,2}^{\rm\,lin} (n_2)
      = \frac{1 - b_{22} n_2}{\tilde{b}_{12}}\, ~.
\end{eqnarray}
For given $n_2$\,, therefore, the domain of $p^{\rm\,lin}$ (\ref{eq:p-lin}) is
\begin{equation}
\label{eq:avr_lin}
  0 \le n_1^{\rm\,lin} < \min \l\{ \hat{n}_{1,1}^{\rm\,lin} (n_2)\,,
                                       ~\hat{n}_{1,2}^{\rm\,lin}(n_2) \r\} ~.
\end{equation}

In the non-linear approximation there is solely one limiting density
\begin{eqnarray}
  \label{eq:n1xnl}
  \hat{n}_1^{\rm\,nl} (n_2)
    &=& {\textstyle
          \frac{1 - 2\,b_{12} n_2
                + \sqrt{(1-2b_{12} n_2)^2 + 4b_{11} n_2 \,(1-b_{22} n_2)}}
               {2\,b_{11}} }\, ~.
\end{eqnarray}
Consequently, for given $n_2$ the domain of $p^{\rm\,nl}$ (\ref{eq:p-nl}) is
\begin{equation}
\label{eq:avr_nl}
  0 \le n_1^{\rm\,nl} < \hat{n}_1^{\rm\,nl} (n_2) ~.
\end{equation}
In the non-linear approximation there is furthermore a region where
the pressure has negative partial derivatives with respect to the smaller
particles' number, $\pd p^{\rm\,nl}/\pd N_2<0$\,.
The condition $\pd p^{\rm\,nl}/\pd N_2=0$ defines the boundary
of this region
\begin{eqnarray}
  \label{eq:n1nl_bd}
  \lefteqn{\hat{n}_1^{\rm\,nl,\,bd} (n_2)}  \\
  &\quad =&
    {\textstyle
      \frac{1 - 2\,(b_{12}-b_{22})\,n_2
            + \sqrt{(1 - 2\,(b_{12}-b_{22})\,n_2)^2 + 8\,(b_{11}-b_{12})\,n_2}}
           {4\,(b_{11}-b_{12})} }\, ~.
    \nonumber
\end{eqnarray}
For given $n_2$ a negative derivative $\pd p^{\rm\,nl}/\pd N_2<0$
occurs only at a density $n_1>\hat{n}_1^{\rm\,nl,\,bd}$\,, while
the derivative $\pd p^{\rm\,nl}/\pd N_1$ is always positive
for $R_2\le R_1$ as readily checked.

In Fig.~\ref{figs:1}\,(a) the functions $\hat{n}_1(n_2)$
are presented in {\em dimensionless\/} variables $\hat{n}_1 b_{11}$
and $n_2 b_{22}$\,.
The properties of these dimensionless functions depend only on
the ratio of the two radii $R_2/R_1$\,.
The smaller this ratio is, the higher is the maximum value of
$\hat{n}_1^{\rm\,nl}$\,, while the region of negative derivatives
$\pd p^{\rm\,nl}/ \pd N_2$ becomes narrower.
The straight line $\hat{n}_{1,1}^{\rm\,lin} (n_2)$ starts
always at $1/b_{11}$\,, but its slope decreases for smaller $R_2/R_1$\,,
whereas $\hat{n}_{1,2}^{\rm\,lin}(n_2)$ ends at $1/b_{22}$ and its
slope increases.
The pressure of the {\em separated model\/} (\ref{eq:p-sp_CE}) would
yield one straight line from $n_1 b_{11}=1$ to $n_2 b_{22}=1$
in Fig.~\ref{figs:1}\,(a)\,, for any ratio of the radii $R_1$ and $R_2$\,.

For very small ratios $R_2/R_1$\,, i.\,e.~for $R_2\to 0$\,,
one finds from Eq.~(\ref{eq:v-nl}) that
$v^{\rm\,nl}\to N_1 b_{11}\,[1-\frac{3}{4} N_2/(N_1+N_2)]$\,.
This yields the maximum density $\max(\hat{n}_1^{\rm\,nl})=4/b_{11}$
for $N_2 \gg N_1$\,.
Thus $\hat{n}_1^{\rm\,nl}$ exceeds the maximum density of the linear
approximation or of the corresponding one-component {\sl VdW\/} gas,
$\max(\hat{n}_{1,1}^{\rm\,lin})=\max(n_1^{\rm\,oc})=1/b_{11}$\,, by a
factor of four in this case.

Note that the value $4/b_{11}$ appears in the linear approximation as well:
For $v_2^{\rm\,lin}\to V$ it is $\max(\hat{n}_{1,2}^{\rm\,lin})=4/b_{11}$
at $n_2=0$\,, but this density cannot be achieved because
$p_1^{\rm\,lin}(n_1,n_2)$ is infinite for $n_1\ge 1/b_{11}$\,.

Let us consider now the consequences of negative derivatives
$\pd p^{\rm\,nl}/\pd N_2<0$ in the non-linear approximation.
If a negative $\pd p^{\rm\,nl}/\pd N_2$ occurs for a density
$n_1^{\,\prime}={\rm const.}$ at $n_2=0$\,, the pressure
$p^{\rm\,nl}(n_1^{\,\prime},n_2)$ has a minimum at a certain density
$n_{2,\rm\,min}>0$\,, which is determined by the boundary
$\hat{n}_1^{\rm\,nl,\,bd}(n_2)$\,.
For increasing $n_1$ along the boundary, 
consequently, the non-linear pressure is always lower
than for increasing $n_1$ at fixed $n_2=0$\,.
Hence along the boundary higher densities can be achieved,
in particular $n_1>1/b_{11}$\,.

Therefore, exceeding of $n_1^{\rm\,nl}=1/b_{11}$ requires that
the boundary starts inside the the non-linear domain at $n_2=0$\,.
Thus the condition $\hat{n}_1^{\rm\,nl,\,bd}(0)<1/b_{11}$ provides
the critical radius $R_{2,\rm\,crit}$ (\ref{eq:R2crit_CE})\,,
\begin{equation}
  \label{eq:R2crit_appd}
  b_{11}<2 b_{12} \quad \leadsto \quad
  R_{2,\rm\,crit}(R_1) = (\sqrt[3]{4} - 1)\,R_1  ~.
\end{equation}
On the other hand the boundary starts at $n_1=8/(14\,b_{11})$ for $R_2\to 0$
at $n_2=0$\,, i.\,e.~for any density $b_{11} n_1\le 8/14\approx 1/2$
negative values of $\pd p^{\rm\,nl}/\pd N_2$ do not occur for any radii.

Although it is pathological that for {\em high\/} densities $n_1$
the non-linear pressure firstly decreases, if particles of the second
and smaller component are added to the system, there is a reasonable
explanation for the lowered pressure along the boundary (\ref{eq:n1nl_bd})
for small radii $R_2<R_{2,\rm\,crit}$\,.

Consider, for instance, $n_1 b_{11}=0.9$ in Fig.~\ref{figs:1}\,(a)\,.
Since it is $R_2/R_1=0.4$\,, the dimensionless density of the small
particles at the boundary $\hat{n}_1^{\rm\,nl,\,bd}(n_2)$ nearly
vanishes, $n_2 b_{22}\approx 0.05$\,, whereas the absolute amounts
of the small and large particles are about equal.
For the excluded volume interaction of the large particles in the
pressure formula (\ref{eq:p-nl})\,, therefore, the influence of the
mixed term $b_{12}$ becomes comparable to that of the distinctly larger
non-mixed term $b_{11}$\,.

For $R_2/R_1\ll 1$ one obtains $n_2\gg n_1$ near the boundary
at $n_1 b_{11}=0.9$\,, i.\,e.~the large particles are completely surrounded
by the smaller particles and interact mostly with these but hardly with
other large particles anymore.
In this situation, consequently, the excluded volume interaction of the
large particles is governed by the essentially smaller $b_{12}$\, and not by
$b_{11}\le 8\,b_{12}$\,.

One might interprete this behaviour as an effective attraction
between small and large particles, but it is rather a strong reduction
of the large particles' excluded volume suppression.

%

As {\sl VdW\/} approximations are low density approximations
they coincide for these densities, but they evidently become inadequate
near the limiting densities:
both discussed formulations do evidently not match the real
{\em gas of rigid spheres\/} there.

For high densities the linear approximation behaves natural,
i.\,e.~it is $\pd p^{\rm\,lin}/\pd N_q>0$ always.
However, one has to introduce the additional terms
$\tilde{b}_{12}$ and $\tilde{b}_{21}$\,.
For the choice (\ref{eq:bTi-s}) these terms provide a one-component-like
behaviour in the limits $R_2=R_1$ and $R_2=0$\,, but they have no concrete
physical meaning.

In the non-linear approximation there occur pathologic pressure derivatives
$\pd p^{\rm\,nl}/\pd N_2<0$ for $R_2\ll R_1$\,.
However, the non-linear formulae may be used for special purposes,
e.\,g.~for $n_1>1/b_{11}$ at intermediate $n_2 b_{22}$\,, where the linear
approximation is yet invalid.


\section{Stability of the Non-linear Approximation}
\label{nl_appd}

The non-linear enhancement in the GCE or the occurence of
negative values for $\pd p^{\rm\,nl}/\pd N_q$ in the CE
suggest a further investigation concerning the thermodynamical
stability of the non-linear approximation.

One can readily check that in the CE it is $\pd p^{\rm\,nl}/\pd V<0$
generally, so there is no {\em mechanical\/} instability.
To investigate wether there is a {\em chemical\/} instability \cite{Reichl}
it is necessary to study partial derivatives with respect to the
particle numbers, $\pd/\pd N_q$\,, of the {\em chemical potentials\/}
\begin{eqnarray}
  \mu_p(T,V,N_1,N_2)
    &\equiv& - T\,{\textstyle \frac{\pd}{\pd N_p} }\,\ln[Z(T,V,N_1,N_2)] ~.
\end{eqnarray}
Partial derivatives of the {\em pressure\/} with respect to the
particle numbers $\pd p/\pd N_q$ have no relevance here.

For the examination of chemical stability it is appropriate
to switch from the {\em free energy\/} of the CE,
$F(T,V,N_1,N_2) \equiv - T \,\ln[Z(T,V,N_1,N_2)]$\,,
to the {\sl Gibbs\/} {\em free energy\/} or {\em free enthalpie\/}
\begin{eqnarray}
  G(T,p,N_1,N_2) &\equiv& F + p\,V = \mu_1 N_1 + \mu_2 N_2 ~,
\end{eqnarray}
where $\mu_q(T,p,N_1,N_2) \equiv \pd G / \pd N_q$\,.
This requires that $p(T,V,N_1,N_2)$ can be solved for $V(T,p,N_1,N_2)$\,,
which is the case for the non-linear approximation,
\begin{eqnarray*}
  V^{\rm\,nl}(T,p,N_1,N_2)
    &=& {\textstyle \frac{N_1+N_2}{p/T} } + N_1 b_{11}
        + N_2 b_{22} - {\textstyle \frac{N_1\,N_2}{N_1+N_2} }D\,.
\end{eqnarray*}
Further it is useful to introduce the molar free enthalpie
$g \equiv G/(N_1+N_2) = g(T,p,x_1)$ with the molar fractions
$x_1 \equiv N_1/(N_1+N_2)$ and $(1-x_1) = x_2 \equiv N_2/(N_1+N_2)$ of
component 1 and 2\,, respectively.
Then the chemical stability of a binary mixture \cite{Reichl}
corresponds to the condition
\begin{equation}
  \label{eq:ch-stab}
  {\textstyle \frac{\pd^2}{\pd x_1^{\,2}} } \,g(T,p,x_1)
    = {\textstyle \frac{\pd \mu_1(T,p,x_1)}{\pd x_1}
                  - \frac{\pd \mu_2(T,p,x_1)}{\pd x_1} }
    > 0 ~.
\end{equation}
For the non-linear approximation one obtains
\begin{eqnarray}
  \lefteqn{g^{\rm\,nl}(T,p,x_1)}  \nonumber \\
    &\quad =&
      x_1 \l\{ T\,\ln\l[{\textstyle \frac{x_1}{\phi_1} \frac{p}{T} }\r]
               + p \l( b_{11} - (1-x_1)^2\,D \r) \r\}  \\
    &  &
      {}+ (1-x_1) \l\{ T\,\ln\l[{\textstyle \frac{1-x_1}{\phi_2}
                                            \frac{p}{T} }\r]
                       + p \l( b_{22} - x_1^{\,2}\,D \r) \r\} ~,  \nonumber
\end{eqnarray}
and thus condition (\ref{eq:ch-stab}) is satisfied:
\begin{eqnarray}
  {\textstyle \frac{\pd^2}{\pd x_1^{\,2}} } \,g^{\rm\,nl}(T,p,x_1)
    = {\textstyle \frac{T}{x_1} }
      + {\textstyle \frac{T}{1-x_1} } + p\,2D
    > 0 ~.
\end{eqnarray}
Therefore, the system described by the non-linear approximation is
thermodynamically stable -- despite the pathologic behaviour in special cases.
Due to the equivalence of the thermodynamical ensembles this is true for any
representation of the model.


\section{The Two VdW Approximations in the GCE}
\label{GCE_appd}

In this part we will study the non-linear and the linear
approximation in the grand canonical ensemble.
As in the CE the differences between the linear and non-linear
approximation occur only for high densities of the larger particles,
$n_q$\,, which correspond to large chemical potentials $\mu_q$ in the
grand canonical treatment.
Therefore, we will study the limit given by Eq.~(\ref{eq:mu1lim})\,:
$\mu_1/T \to \infty$ for constant $T, \mu_2$ and $R_2 \le R_1$\,,
where it is $\xi_1 \to \infty$\,.

The {\em transcendental exponents\/} of both approximations contain an
{\em attractive\/} part besides the usual {\sl VdW}-like {\em suppressive\/} part
$\exp[-(p/T)\,b_{qq}]$\,:
In the linear approximation (\ref{eq:xi1-lin}, \ref{eq:xi2-lin}) the
suppression is reduced, but the transcendental exponents are always negative
-- whereas the suppression in the non-linear approximation
(\ref{eq:xi1}, \ref{eq:xi2}) is not only reduced, but the transcendental
exponent of $\xi_2^{\rm\,nl}$ can even become positive in the above limit.

To examine the latter we rewrite the coupled transcendental
equations (\ref{eq:xi1}, \ref{eq:xi2}) as
\begin{eqnarray}
  \label{eq:xi1_A}
  \xi_1^{\rm\,nl}
    &=& \phi_1 \, \exp \l[{\textstyle \frac{\mu_1}{T} }\r] \times  \\
    & & {} \times {\textstyle
        \exp \l[-\l(\xi_1^{\rm\,nl}+\xi_2^{\rm\,nl}\r) b_{11} \,
                 \l(1 - \frac{D/b_{11}}
                             {\l(\xi_1^{\rm\,nl}/\xi_2^{\rm\,nl} + 1\r)^2}\r)
                \r] } ~,
        \nonumber \\
  \label{eq:xi2_A}
  \xi_2^{\rm\,nl}
    &=& \phi_2 \, \exp \l[{\textstyle \frac{\mu_2}{T} }\r] \times  \\
    & & {} \times {\textstyle
        \exp \l[-\l(\xi_1^{\rm\,nl} + \xi_2^{\rm\,nl}\r) b_{22} \,
                 \l(1 - \frac{D/b_{22}}
                             {\l(1 + \xi_2^{\rm\,nl}/\xi_1^{\rm\,nl}\r)^2}\r)
                \r] } ~.  \nonumber
\end{eqnarray}
If $R_2$ is sufficiently smaller than $R_1$\,,
then $D/b_{22}$ is larger than unity and the transcendental exponent of
$\xi_2^{\rm\,nl}$ (\ref{eq:xi2_A}) can become positive,
\begin{eqnarray}
  0 < {\textstyle \frac{D}{b_{22}}
      - \l( 1 + \frac{\xi_2^{\rm\,nl}}{\xi_1^{\rm\,nl}} \r)^2}  \nonumber \\
  \label{eq:a-Def}
  \Longleftrightarrow
    &~& {\textstyle \frac{\xi_2^{\rm\,nl}}{\xi_1^{\rm\,nl}}
         < a_2 \equiv \sqrt{\frac{D}{b_{22}}} - 1 } ~.
\end{eqnarray}
The coefficient $a_2=a_2(R_1/R_2)$ introduced here is the crucial
combination of excluded volumes in the non-linear approximation.
It characterises the behaviour of this approximation in the
limit (\ref{eq:mu1lim})\,, i.\,e.~for $\xi_1\to\infty$\,.

For equal radii $R_2=R_1$ it is $a_2=-1$\,, and one has full {\sl VdW}-like
suppression, $\propto \exp[-(p/T)\,b_{qq}]$\,.
Negative $a_2$ indicate the strength of suppression of $\xi_2^{\rm\,nl}$
for increasing $\xi_1^{\rm\,nl}$\,.
For $-1<a_2<0$ the suppression is reduced, most strongly for $a_2\approx 0$\,.
In the case $a_2=0$ there is no suppression in the limit (\ref{eq:mu1lim})
but $\xi_2^{\rm\,nl}\to A_2={\rm const.}$~for $\xi_1^{\rm\,nl}\to\infty$\,.
Thus, the condition $a_2=0$ provides the critical radius,
\begin{equation}
  \label{eq:R2crit_GCE}
  D/b_{22} = 1
    \quad \Longleftrightarrow \quad
  R_{2,\rm\,crit} = (\sqrt[3]{4}-1)\, R_1 ~,
\end{equation}
which coincides with the canonical result (\ref{eq:R2crit_appd})\,.

For positive $a_2$ or $R_2<R_{2,\rm\,crit}$ the {\em non-linear enhancement\/}
occurs:
$\xi_2^{\rm\,nl}$ is enhanced by increasing $\xi_1^{\rm\,nl}$
as long as the second exponent in (\ref{eq:xi2_A}) is positive,
i.\,e.~for $\xi_2^{\rm\,nl}<a_2\,\xi_1^{\rm\,nl}$\,.
The transcendental factor of $\xi_1^{\rm\,nl}$ (\ref{eq:xi1_A})
has only a reduced suppression in this case.
According to Eq.~(\ref{eq:a-Def}) one obtains
\begin{equation}
  \label{eq:xi1lim}
  \xi_2^{\rm\,nl} \to a_2 \, \xi_1^{\rm\,nl}~, \qquad
  \mbox{but also} \qquad
  n_2^{\rm\,nl} \to a_2 \, n_1^{\rm\,nl}~,
\end{equation}
since it is $n_2^{\rm\,nl}/n_1^{\rm\,nl}=\xi_2^{\rm\,nl}/\xi_1^{\rm\,nl}$
due to Eqs.~(\ref{eq:n1}, \ref{eq:n2})\,.

Using Eqs.~(\ref{eq:xi1lim}) and (\ref{eq:a-Def}) one obtains for
the particle densities (\ref{eq:n1}, \ref{eq:n2})
\begin{eqnarray}
  n_1^{\rm\,nl}
    &\to& {\textstyle \frac{1}{b_{11} - a_2^{\,2}\,b_{22}} }
            = {\textstyle \frac{1}{b_{11}
               - (\sqrt{D}-\sqrt{b_{22}}\,)^2} } \,~,  \\
  n_2^{\rm\,nl}
    &\to& a_2 \, n_1^{\rm\,nl}  \nonumber \\
\label{eq:n2-nl_lim}
    &\to& {\textstyle \frac{a_2}{b_{11} - a_2^{\,2}\,b_{22}} }
            = {\textstyle \frac{\sqrt{b_{22}}}{b_{22}}
              \, \frac{\sqrt{D} - \sqrt{b_{22}}}
                      {b_{11} - (\sqrt{D} - \sqrt{b_{22}}\,)^2} } \,~.
\end{eqnarray}
It is clearly seen that the density $n_1^{\rm\,nl}$ can exceed $1/b_{11}$
for positive $a_2$\,.
As in the CE the maximum value, $\max(n_1^{\rm\,nl})=4/b_{11}$\,,
is achieved for $R_2=0$ or $a_2=\infty$\,.
Then, in the limit (\ref{eq:mu1lim}) the density of the second component
diverges, $n_2^{\rm\,nl}\to\infty$\,,
but its total excluded volume vanishes, $n_2^{\rm\,nl}\,b_{22}\to 0$\,,
as seen from Eq.~(\ref{eq:n2-nl_lim})\,.

There is yet another case, where the condition (\ref{eq:a-Def}) is
fulfilled, the {\em early enhancement\/}\,:
$\xi_2^{\rm\,nl}$ can be enhanced with increasing $\mu_2$ for constant
$T$ and $\mu_1$\,, if $\mu_1$ is sufficiently large.
This enhancement takes place only at small $\mu_2$\,, and it obviously
vanishes when $\mu_2$ is large enough so that
$\xi_2^{\rm\,nl}\ge a_2\xi_1^{\rm\,nl}$\,.
The early enhancement is the direct analogue to a negative derivative
$\pd p^{\rm\,nl}/\pd N_q<0$ in the CE\,.

The coupled transcendental equations of the linear approximation
(\ref{eq:xi1-lin}, \ref{eq:xi2-lin}) may be rewritten similarly to
Eqs.~(\ref{eq:xi1_A}, \ref{eq:xi2_A})\,.
For the choice (\ref{eq:bTi-s}) one obtains in terms of $D$ from
Eq.~(\ref{eq:Def_D})
\begin{eqnarray}
  \label{eq:xi1-lin_D}
  \lefteqn{\xi_1^{\rm\,lin} (T, \mu_1, \mu_2)}  \\
  &\quad =&
    A_1 ~ \exp \l[{\textstyle -\l(\xi_1^{\rm\,lin}+\xi_2^{\rm\,lin}\r) b_{11}
                  \l(1 - \frac{\xi_2^{\rm\,lin}}
                              {\xi_1^{\rm\,lin}+\xi_2^{\rm\,lin}}
                         \frac{D}{b_{11}+b_{22}}\r) }\r] ~,
    \nonumber \\
  \label{eq:xi2-lin_D}
  \lefteqn{\xi_2^{\rm\,lin} (T, \mu_1, \mu_2)}  \\
  &\quad =&
    A_2 ~ \exp \l[{\textstyle -\l(\xi_1^{\rm\,lin}+\xi_2^{\rm\,lin}\r) b_{22}
                  \l(1 - \frac{\xi_1^{\rm\,lin}}
                              {\xi_1^{\rm\,lin}+\xi_2^{\rm\,lin}}
                         \frac{D}{b_{11}+b_{22}}\r) }\r] ~.
    \nonumber
\end{eqnarray}
In this case the condition for the enhancement of $\xi_2^{\rm\,lin}$
for $R_2\le R_1$ would be
\begin{eqnarray}
  {\textstyle \frac{\xi_2^{\rm\,lin}}{\xi_1^{\rm\,lin}}
    < \tilde{a}_2 \equiv - \frac{2b_{12}}{b_{11}+b_{22}} } ~.
\end{eqnarray}

As $\tilde{a}_2$ is always negative the {\sl VdW}-like
suppression is only reduced in this approximation.
Like in the non-linear approximation $\tilde{a}_2=-1$ corresponds to
equal radii $R_2=R_1$ and full {\sl VdW}-like suppresion, whereas
$\tilde{a}_2=-1/4$ ($R_2=0$) corresponds to the most strongly reduced
suppresion in the linear approximation.

Thus, the densities $n_q^{\rm\,lin}$ can not exceed the maximum value
$1/b_{qq}$ of the corresponding one-component case.
Furthermore, the density $n_2^{\rm\,lin}$ (\ref{eq:n2-lin}) vanishes
in the analogous limit to (\ref{eq:mu1lim})\,, $\xi_1^{\rm\,lin}\to\infty$\,,
in contrast to the non-linear behaviour.

In the case $R_1\le R_2$ one would evidently investigate the
coefficients $a_1\equiv(\sqrt{D/b_{11}}-1)$ in the non-linear
and $\tilde{a}_1\equiv-(2b_{12})/(b_{11}+b_{22})=\tilde{a}_2$
in the linear approximation, respectively.




\end{document}